\documentclass[aps,prl,showpacs,preprintnumbers,
              twocolumn,nofootinbib,float,superscriptaddress,longbibliography]{revtex4-1}

\usepackage{graphics}      
\usepackage{graphicx}      
\usepackage{longtable}     
\usepackage{url}           
\usepackage{bm}            
\usepackage{amsmath}
\usepackage{amssymb}
\usepackage{float}
\usepackage{physics}
\usepackage{amsmath,amssymb}
\usepackage[colorlinks=true, pdfstartview=FitV, linkcolor=red, citecolor=blue, urlcolor=blue]{hyperref}
\usepackage{color}
\usepackage[utf8]{inputenc}

\newcommand{\der}{\mathrm{d}}

\newcommand{\as}{\alpha_\mathrm{s}}

\usepackage{capt-of}

\begin{document}

\title{Multi-gluon correlations and evidence of saturation from dijet measurements\\ at an Electron Ion Collider}

\author{Heikki Mäntysaari}
\email{heikki.mantysaari@jyu.fi  }
\affiliation{\small Department of Physics, University of Jyv\"askyl\"a,
P.O. Box 35, 40014 University of Jyv\"asky\"a, Finland}
\affiliation{\small Helsinki Institute of Physics, P.O. Box 64, 00014 University of Helsinki, Finland}

\author{Niklas Mueller}
\email{nmueller@bnl.gov }
\affiliation{\small Physics Department, Brookhaven National Laboratory, Bldg. 510A, Upton, NY 11973, USA}

\author{Farid Salazar}
\email{farid.salazarwong@stonybrook.edu}
\affiliation{\small Physics Department, Brookhaven National Laboratory, Bldg. 510A, Upton, NY 11973, USA}
\affiliation{\small Department of Physics and Astronomy, Stony Brook University, Stony Brook, New York 11794, USA }

\author{Bj\"orn Schenke}
\email{bschenke@bnl.gov}
\affiliation{\small Physics Department, Brookhaven National Laboratory, Bldg. 510A, Upton, NY 11973, USA}

\date{\today}
\begin{abstract}
We study inclusive and diffractive dijet production in electron-proton and electron-nucleus collisions within the Color Glass Condensate effective field theory. We compute dijet cross sections differentially in both mean dijet transverse momentum $\bm{P}$ and recoil momentum $\bm{\Delta}$, as well as the anisotropy in the relative angle between $\bm{P}$ and $\bm{\Delta}$. We use the nonlinear Gaussian approximation to compute multiparticle correlators for general small $x$ kinematics, employing running coupling Balitsky-Kovchegov evolution to determine the dipole amplitude at small $x$. Our results cover a much larger kinematic range than accessible in previous computations performed in the correlation limit approximation, where it is assumed that $|\bm{P}| \gg |\bm{\Delta}|$. We validate this approximation in its range of applicability and quantify its failure for $|\bm{P}| \lesssim |\bm{\Delta}|$.
We also predict significant target-dependent deviations from the correlation limit approximation for $|\bm{P}| > |\bm{\Delta}|$ and $|\bm{P}| \lesssim Q_s$, which offers a straightforward test of gluon saturation and access to multi-gluon distributions at a future electron  ion collider. 
\end{abstract}
\maketitle

{\bf Introduction.}
To gain a complete understanding of the complex multi-parton structure of nuclei at small $x$, measurements of a multitude of processes in high energy $e+p(A)$ collisions over a wide range of kinematics are necessary. A future electron ion collider (EIC) \cite{Boer:2011fh,Accardi:2012qut,Aschenauer:2017jsk} will provide an ideal tool for such an endeavor, with dijet production being one of the most important processes to access the structure of gluon fields and their non-linear dynamics inside protons and heavier nuclei.

While coherent diffractive dijet production allows to access the target's spatial geometry \cite{Altinoluk:2015dpi,Hatta:2016dxp,Mantysaari:2019csc,Salazar:2019ncp,Hatta:2019ixj}, inclusive and incoherent diffractive dijet cross sections are sensitive to multi-gluon correlations in the target \cite{Dominguez:2011wm,Mueller:2013wwa} (see also~\cite{Mantysaari:2016ykx,Mantysaari:2016jaz,Mantysaari:2017dwh,Mantysaari:2018zdd,Mantysaari:2019jhh}). In the near back-to-back correlation limit, where the mean dijet momentum is much larger than the recoil momentum, the inclusive dijet production cross section can be expressed in terms of the Weizs\"acker-Williams transverse momentum dependent gluon distributions (TMDs), allowing experimental access to these fundamental quantities \cite{Dominguez:2011wm,Dominguez:2011br,Metz:2011wb,Dumitru:2015gaa,Dumitru:2018kuw}. 
We advocate going beyond this limit to allow for deeper insights into the multi-gluon structure of the nucleus. Inclusive and diffractive (incoherent) dijets are sensitive to the quadrupole and dipole-dipole correlators of light-like Wilson lines, respectively. These are among the fundamental objects describing the gluon structure at small $x$.

We present the first evaluation of inclusive and incoherent diffractive dijet cross sections and their azimuthal anisotropies for general small-$x$ kinematics in the Color Glass Condensate (CGC) effective field theory (EFT) at leading order in $\alpha_s$, resumming all terms $\sim\as \ln 1/x$. For inclusive dijets, our results explicitly validate the correlation limit approximations in the kinematic region $|\bm{P}|\gg|\bm{\Delta}|$ and 
extend our knowledge of dijet production to the region $|\bm{P}|\lesssim|\bm{\Delta}|$, where deviations from the correlation limit turn out to be large. We further show that corrections to the correlation limit approximation also become important when $|\bm{P}|\lesssim Q_s$, even when $|\bm{P}|> |\bm{\Delta}|$ holds. These corrections, enhanced by the saturation scale $Q_s$, probe genuine multi-gluon correlations \cite{Dumitru:2016jku,Altinoluk:2019wyu}, and are not encompassed by the resummed kinematic twists of the improved TMD framework \cite{Altinoluk:2019fui} (see also \cite{Marquet:2007vb,Lappi:2012nh,Iancu:2013dta,Kotko:2015ura,vanHameren:2016ftb,Albacete:2018ruq} for forward dijets in dilute-dense hadronic collisions, and experimental measurements from RHIC~\cite{Adare:2011sc,Braidot:2011zj}).

Calculations of the elliptic anisotropy employing multi-gluon correlators deviate strongly from the correlation limit for $|\bm{P}|\lesssim|\bm{\Delta}|$. In particular, for transverse polarization the calculated elliptic modulation is qualitatively different from that in the correlation limit, as a maximum appears both as a function of $|\bm{P}|$ and $|\bm{\Delta}|$.

For the first time within the CGC EFT (see also~\cite{Aktas:2007hn,Guzey:2016awf,Helenius:2019gbd}), we predict the incoherent diffractive cross section, the dominant component of the total diffractive cross section for $|\bm{\Delta}| \gtrsim 0.2 $ GeV in the case of a large nucleus. Our calculation predicts characteristic features of the cross section's elliptic anisotropy as a function of $|\bm{P}|$ and $|\bm{\Delta}|$, involving sign changes and minima, which should be observable experimentally.

We compute the fraction of diffractive dijet events as a function of the mean dijet momentum. It increases with the mass number of the nucleus and decreases with $Q^2$ at a slower rate than expected in the small dipole expansion, signaling gluon saturation~\cite{Kowalski:2008sa}. 

{\bf Dijet production in high energy DIS.}
In the dipole picture of high energy deeply inelastic scattering (DIS), the production of a forward $q\bar{q}$ dijet can be seen as the splitting of a virtual photon $\gamma^*$ into a quark-antiquark dipole and its subsequent eikonal scattering off the target's color field. We work in a frame in which the virtual photon and nucleon in the target have zero transverse momenta\footnote{We denote 2D transverse vectors as $\bm{x}$, with magnitude $|\bm{x}|$. }. The photon has virtuality $Q^2$ and four momentum $q^\mu = (-Q^2/2q^-,q^-,\bm{0} )$. Neglecting its mass, the nucleon has energy $E_n$ and four momentum $P^\mu_n = (\sqrt{2}E_n,0,\bm{0})$. The center of mass energy of the virtual photon-nucleon system is $W$. The transverse momenta of the outgoing quark and antiquark are $\bm{p}_1$ and $\bm{p}_2$, their longitudinal momentum fractions are $z_1$ and $z_2$, with $z_{i} = p_i^-/q^- = 2 E_n |\bm{p}_{i}| e^{-y_i}/W^2$, where $p_i^-$ and $y_i$ are the quark and antiquark longitudinal momenta and rapidities in this frame, respectively.

Expressed using the momenta
$\bm{P} = z_2 \bm{p}_1 - z_1 \bm{p}_2$ and $
\bm{\Delta} = \bm{p}_1+\bm{p}_2$, at leading order in $\alpha_s$, the cross sections for dijet production of massless quarks for longitudinal ($L$) and transverse ($T$) photon polarization read \cite{Dominguez:2011wm, Dominguez:2011br,Roy:2018jxq}
\begin{widetext}
\vspace{-0.4cm}
\begin{align}
 \frac{\der \sigma^{\gamma^* A \rightarrow q\bar{q} X} _{L}}{\der y_1 \der y_2 \der^2 \bm{P} \der^2 \bm{\Delta}} = &\frac{8  \alpha_{e} Z^2_f N_c S_\perp}{(2 \pi)^6}  \delta_z z^3_1 z^3_2 Q^2 \int \limits_{\substack{\bm{b}-\bm{b}' \\\bm{r},\bm{r}'}} e^{-i \bm{P} \cdot (\bm{r} -\bm{r}')} e^{-i \bm{\Delta} \cdot (\bm{b} -\bm{b}')} \mathcal{O}_{\bm{r},\bm{b};\bm{r}',\bm{b}'} K_0(\varepsilon_f |\bm{r}|) K_0(\varepsilon_f |\bm{r}'|)\,,
 \label{Full_xsecL}\\
 \frac{\der \sigma^{\gamma^* A \rightarrow q\bar{q} X} _{T}}{\der y_1 \der y_2 \der^2 \bm{P} \der^2 \bm{\Delta}} = & \frac{2 \alpha_{e} Z^2_f N_c S_\perp}{(2 \pi)^6}  \delta_z z_1 z_2 (z_1^2+z_2^2) \varepsilon_f^2  \int \limits_{\substack{\bm{b}-\bm{b}' \\\bm{r},\bm{r}'}} e^{-i \bm{P} \cdot (\bm{r} -\bm{r}')} e^{-i \bm{\Delta} \cdot (\bm{b} -\bm{b}')}  \mathcal{O}_{\bm{r},\bm{b};\bm{r}',\bm{b}'}   \frac{\bm{r}\cdot \bm{r}'}{|\bm{r} ||\bm{r}'|}   K_1(\varepsilon_f |\bm{r}|) K_1(\varepsilon_f |\bm{r}'|)\,.
\label{Full_xsecT}
\end{align}
\vspace{-0.4cm}
\end{widetext}
Here, $\alpha_e=e^2/(4\pi)$ is the electromagnetic coupling, $N_c=3$ is the number of colors, $\delta_z=\delta(1-z_1-z_2)$, $\varepsilon_f^2=z_1 z_2 Q^2$, and $\int_{\bm{x}}=\int \der^2\bm{x}$.
We use $Z_f^2 = (\frac{2}{3})^2+(-\frac{1}{3})^2+(-\frac{1}{3})^2$, corresponding to $u$, $d$ and $s$ quarks. Assuming a homogeneous target, the cross section is proportional to the effective transverse area of the target $S_\perp$. The multi-gluon correlations are encoded in $\mathcal{O}$, defined as
\begin{align}
\mathcal{O}_{\bm{r},\bm{b};\bm{r}',\bm{b}'}^{(4)} &= 1-S^{(2)}_{\bm{x}_1,\bm{x}_2} -S^{(2)}_{\bm{x}'_2,\bm{x}'_1} + S^{(4)}_{\bm{x}_1,\bm{x}_2;\bm{x}'_2,\bm{x}'_1}
\end{align}
for inclusive production, and
\begin{align}
\mathcal{O}_{\bm{r},\bm{b};\bm{r}',\bm{b}'}^{(2,2)} &= 1-S^{(2)}_{\bm{x}_1,\bm{x}_2} -S^{(2)}_{\bm{x}'_2,\bm{x}'_1} + S^{(2,2)}_{\bm{x}_1,\bm{x}_2;\bm{x}'_2,\bm{x}'_1}
\end{align}
for total diffractive (color singlet) production. The $\bm{x}$ coordinates are related to $\bm{r}$ and $\bm{b}$ via $\bm{x}_{1,2} = \bm{b}\pm z_{2,1} \bm{r}$ and $\bm{x}'_{1,2} = \bm{b}'\pm z_{2,1} \bm{r}'$.
The dipole, dipole-dipole, and quadrupole correlators of fundamental light-like Wilson lines $V$ are defined by \cite{Dominguez:2011wm,Lappi:2015vta}
\begin{align}
S^{(2)}_{\bm{x}_1,\bm{x}_2} & = \frac{1}{N_c}  \left \langle  \tr \left( V^\dagger_{\bm{x}_1} V_{\bm{x}_2} \right) \right \rangle \,,\label{Correlator1_xy} \\
S^{(2,2)}_{\bm{x}_1,\bm{x}_2;\bm{x}'_2,\bm{x}'_1} & = \frac{1}{N^2_c}  \left  \langle  \tr \left( V^\dagger_{\bm{x}_1} V_{\bm{x}_2} \right) \tr \left( V^\dagger_{\bm{x}'_2} V_{\bm{x}'_1} \right) \right \rangle \,, \label{Correlator3_xy} \\
S^{(4)}_{\bm{x}_1,\bm{x}_2;\bm{x}'_2,\bm{x}'_1} & = \frac{1}{N_c}  \left  \langle  \tr \left( V^\dagger_{\bm{x}_1} V_{\bm{x}_2} V^\dagger_{\bm{x}'_2} V_{\bm{x}'_1} \right) \right \rangle \,. \label{Correlator2_xy}
\end{align}
where the $\left\langle \cdot \right \rangle$ denote the average over static large $x$ color source configurations in the CGC EFT.
The difference between inclusive and total diffractive processes results solely from the color structures of the correlators.

The correlators $\mathcal{O}_{\bm{r},\bm{b};\bm{r}',\bm{b}'}$ contain both the elastic\footnote{The elastic (coherent) production of dijets is given by Eqs.\, \eqref{Full_xsecL} and \eqref{Full_xsecT} with $\mathcal{O}_{\bm{r},\bm{b};\bm{r}',\bm{b}'} = 1-S^{(2)}_{\bm{x}_1,\bm{x}_2} -S^{(2)}_{\bm{x}'_2,\bm{x}'_1} + S^{(2)}_{\bm{x}_1,\bm{x}_2} S^{(2)}_{\bm{x}'_2,\bm{x}'_1}$.} and inelastic parts. In this work we neglect the impact parameter dependence of the target such that the elastic cross section vanishes at non-zero $\bm{\Delta}$. This amounts to the replacements $\mathcal{O}_{\bm{r},\bm{b};\bm{r}',\bm{b}'}^{(4)} \rightarrow S^{(4)}_{\bm{r},\bm{b};\bm{r}',\bm{b}'}  - S^{(2)}_{\bm{r},\bm{b}} S^{(2)}_{\bm{r}',\bm{b}'}$, and $\mathcal{O}_{\bm{r},\bm{b};\bm{r}',\bm{b}'}^{(2,2)}  \rightarrow S^{(2,2)}_{\bm{r},\bm{b};\bm{r}',\bm{b}'}  - S^{(2)}_{\bm{r},\bm{b}} S^{(2)}_{\bm{r}',\bm{b}'}$,
which restrict the cross sections to the inelastic part and simplify their evaluation.
The correlators above are evaluated at
$x = (Q^2+|\bm{\Delta}|^2+ M^2_{q\bar{q}})/W^2$, 
which follows from kinematics and energy-momentum conservation~\cite{Dumitru:2018kuw,Dominguez:2011wm}, where the invariant mass of the dijet is given by $M_{q\bar{q}}^2=   |\bm{P}|^2/(z_1 z_2)$.

\begin{figure*}[!htb]
    \begin{center}
    \includegraphics[width=18cm]{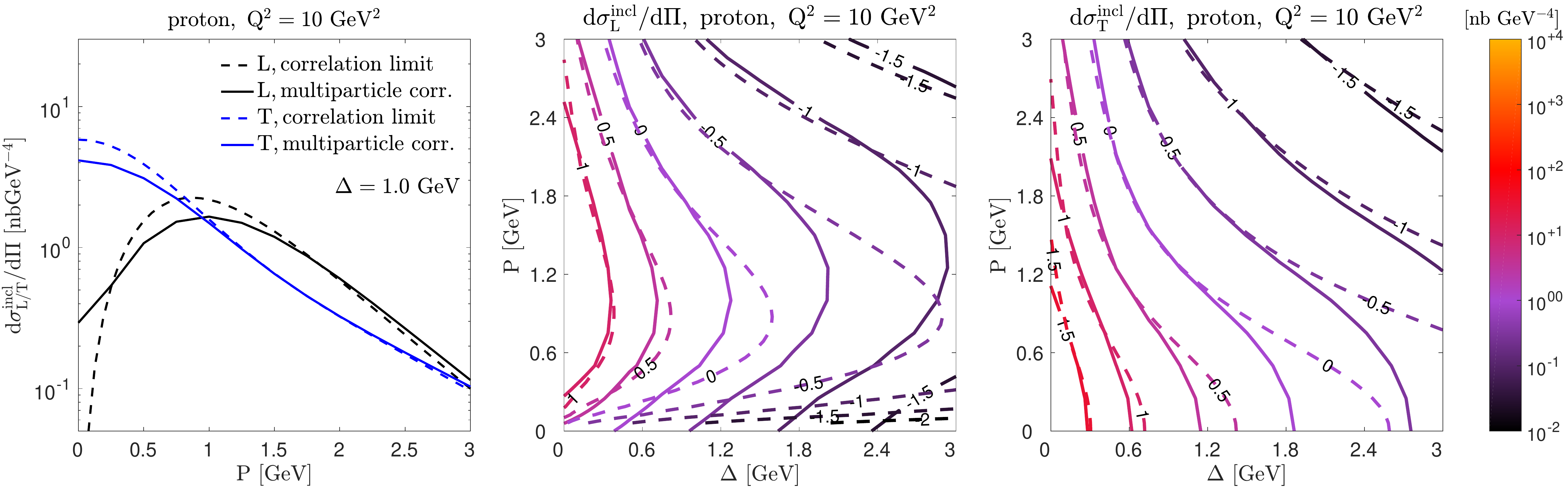}
    \includegraphics[width=18cm]{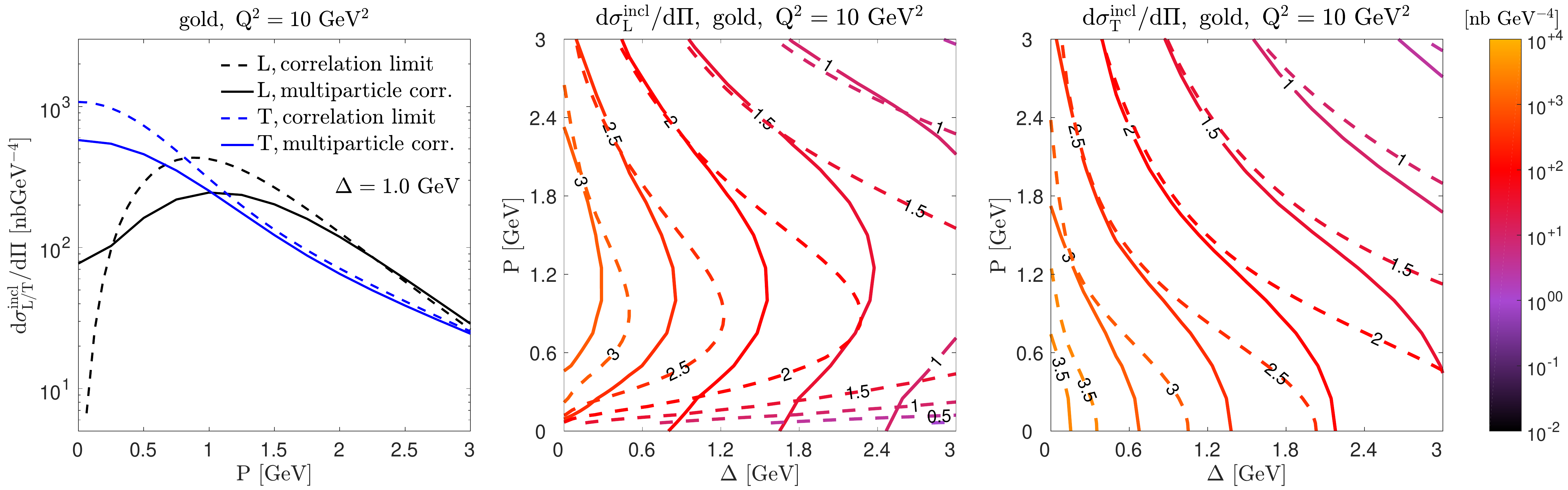}
    \caption{Angle averaged inclusive dijet cross section for proton (upper) and gold (lower) targets. Solid lines: full multiparticle correlator result. Dashed lines: correlation limit approximation. Panels on the left show a vertical section of the contour plots at fixed $|\bm{\Delta}|=1\,{\rm GeV}$. \label{Inclusive_xsec}}
    \end{center}
\end{figure*}

To reduce the computational cost of our calculation, we employ the nonlinear Gaussian approximation, which allows one to express any $n-$point correlator of light-like Wilson lines as a non-linear function of the dipole correlator in Eq.\,\eqref{Correlator1_xy}, and was shown to approximate the full quadrupole operator very well \cite{Dumitru:2011vk}, even after JIMWLK small $x$ evolution for many units in rapidity \cite{JalilianMarian:1996xn,JalilianMarian:1997gr,JalilianMarian:1997jx,Iancu:2001ad,Iancu:2000hn,Ferreiro:2001qy,Iancu:2001md}. The Gaussian approximation yields \cite{Dominguez:2011wm,Lappi:2015vta,Dumitru:2011vk,Dominguez:2008aa,Blaizot:2004wv,Fukushima:2007dy}
\begin{align}
S^{(4)/(2,2)}_{\bm{x}_1,\bm{x}_2;\bm{x}'_2,\bm{x}'_1} &\approx S^{(2)}_{\bm{x}_1,\bm{x}_2} S^{(2)}_{\bm{x}'_2,\bm{x}'_1}\notag\\ & \hspace{-2.6cm}\times \left[ \left( \frac{\sqrt{\Delta} + F_{\bm{x}_1,\bm{x}'_2; \bm{x}_2,\bm{x'}_1}}{2 \sqrt{\Delta}} - \frac{ F_{\bm{x}_1,\bm{x}_2; \bm{x}'_2,\bm{x}'_1}}{N^{(4)/(2,2)}\sqrt{\Delta}} \right) e^{\frac{N_c\sqrt{\Delta}}{4}} \right. \nonumber \\ &\hspace{-2.2cm} +\left.  \left( \frac{\sqrt{\Delta} - F_{\bm{x}_1,\bm{x}'_2; \bm{x}_2,\bm{x'}_1}}{2 \sqrt{\Delta}} + \frac{ F_{\bm{x}_1,\bm{x}_2; \bm{x}'_2,\bm{x}'_1}}{N^{(4)/(2,2)}\sqrt{\Delta}} \right) e^{\frac{-N_c\sqrt{\Delta}}{4}} \right] \nonumber \\ & \hspace{-2.6cm} \ \ \ \ \times e^{-\frac{N_c}{4} F_{\bm{x}_1,\bm{x}'_2; \bm{x}_2, \bm{x}'_1} + \frac{1}{2N_c}F_{\bm{x}_1,\bm{x}_2; \bm{x}'_2, \bm{x}'_1}  } \,,
\end{align}
where the only difference between the two cases is the constant $N^{(4)} = 1$ in case of the quadrupole and $N^{(2,2)} = N_c^2$ in case of the dipole-dipole correlator. We define
\begin{align}
\Delta_{\bm{x}_1,\bm{x}_2; \bm{x}'_2,\bm{x}'_1} &= F^2_{\bm{x}_1,\bm{x}'_2; \bm{x}_2,\bm{x}'_1} + \frac{4}{N_c^2} F_{\bm{x}_1,\bm{x}_2; \bm{x}'_2,\bm{x}'_1} F_{\bm{x}_1,\bm{x}'_1; \bm{x}'_2,\bm{x}_2} \nonumber,\\
F_{\bm{x}_1,\bm{x}_2;\bm{x}'_2,\bm{x}'_1} & = \frac{1}{C_F} \ln\left[\frac{S^{(2)}_{\bm{x}_1,\bm{x}'_2} S^{(2)}_{\bm{x}_2,\bm{x}'_1}}{S^{(2)}_{\bm{x}_1,\bm{x}'_1}S^{(2)}_{\bm{x}_2,\bm{x}'_2}}\right]\notag\,,
\end{align}
with $C_F = (N_c^2-1)/(2N_c)=4/3$ and $\Delta=\Delta_{\bm{x}_1,\bm{x}_2; \bm{x}'_2,\bm{x}'_1}$.

\begin{figure*}[!htb]
    \centering
    \includegraphics[width=18.cm]{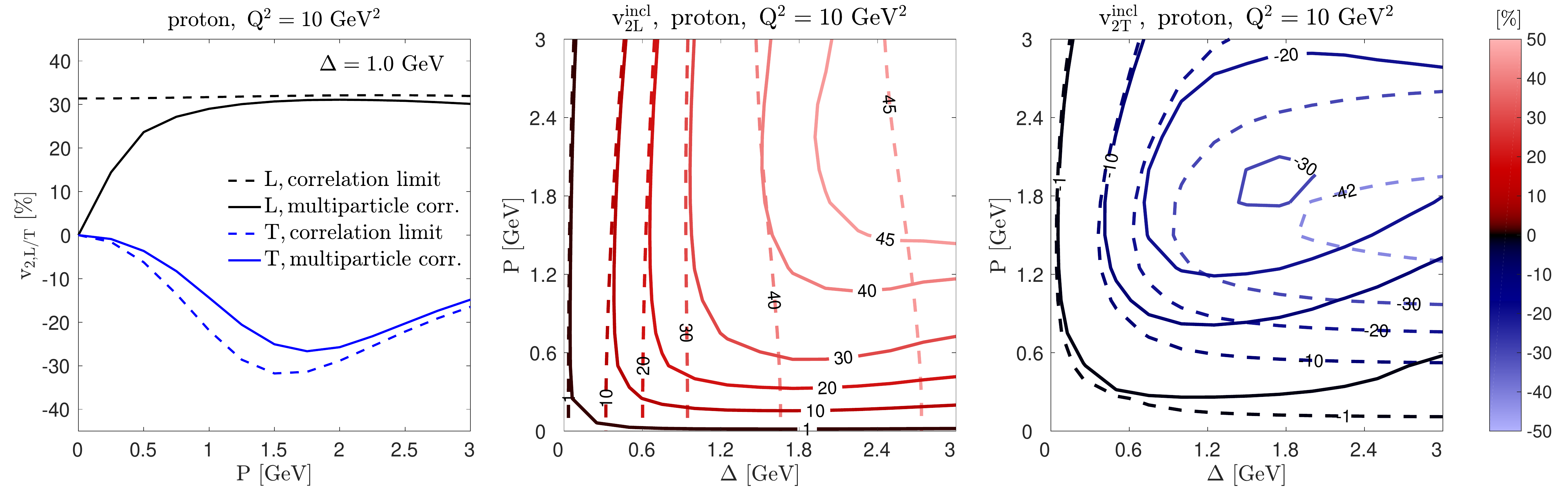}
    \includegraphics[width=18.cm]{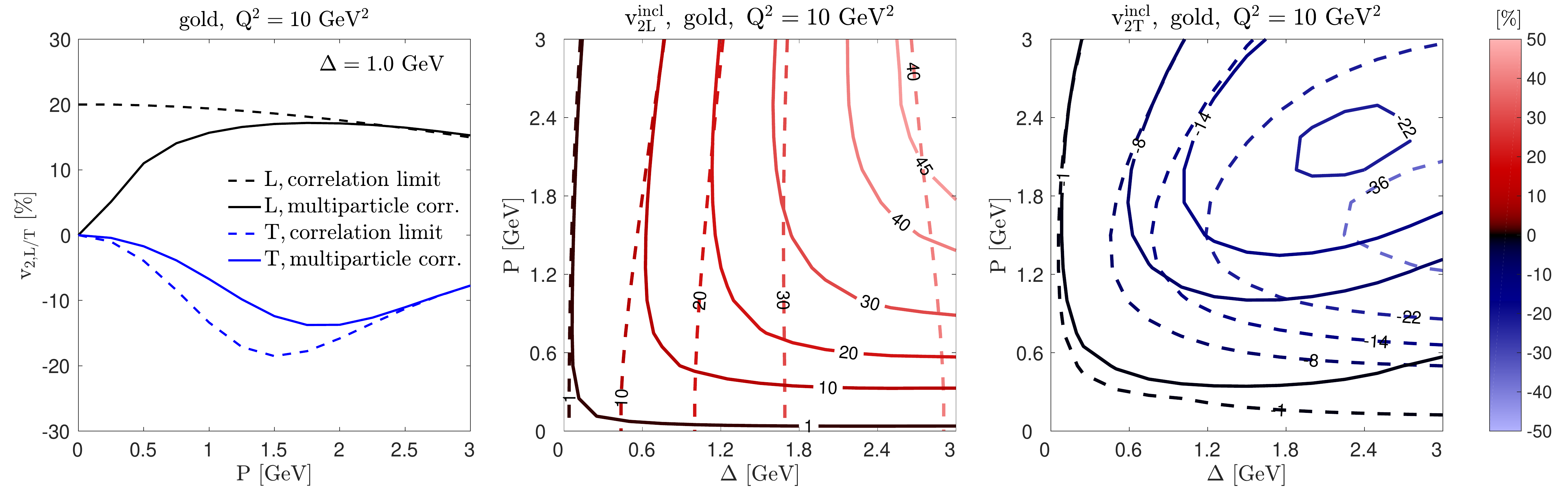}
    \caption{Elliptic anisotropy of inclusive dijet cross sections for proton (upper), and gold (lower). Solid lines: full multiparticle correlator result. Dashed lines: correlation limit approximation. Panels on the left show a vertical section of the contour plots at fixed $|\bm{\Delta}|=1\,{\rm GeV}$. We emphasize the appearance of distinct minima in the $v_{2T}$, which are not captured by the correlation limit approximation. \label{Inclusive_v2}}
\end{figure*}

\begin{figure*}[!htb]
        \centering
    \includegraphics[width=18cm]{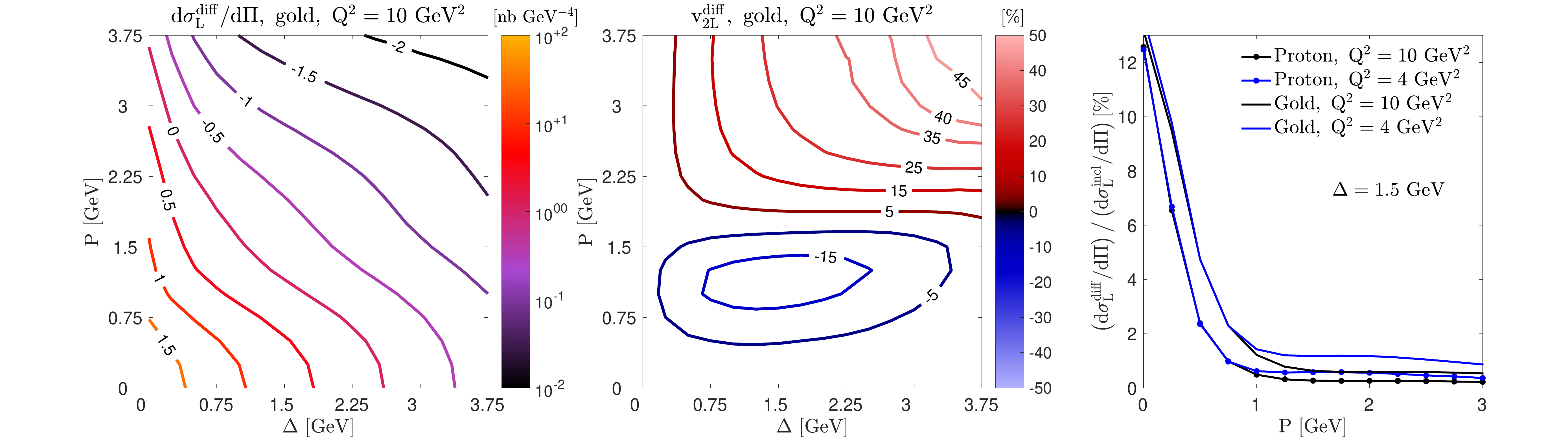}
    \includegraphics[width=18cm]{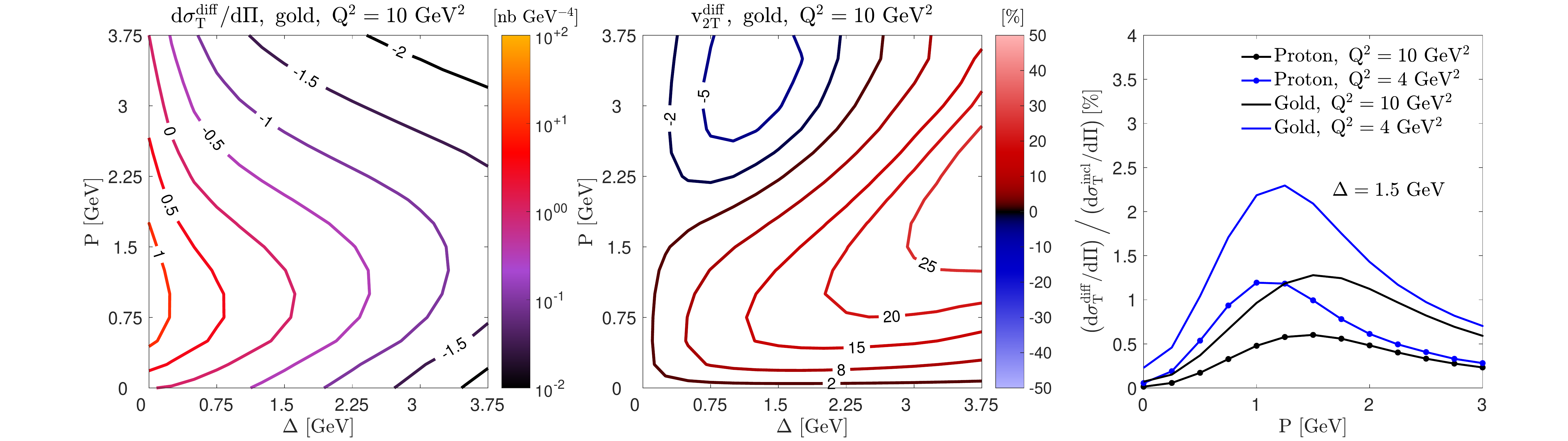}
    \caption{Left: Diffractive angle averaged dijet cross sections. Center: Diffractive elliptic anisotropy. Right: Ratio diffractive to inclusive cross section. Upper panels: Longitudinal. Lower panels: Transverse. 
    \label{Diffractive_xsec}}
\end{figure*}

The dipole correlator $S^{(2)}$ satisfies the leading order Balitsky-Kovchegov (BK) evolution equation~\cite{Balitsky:1995ub,Kovchegov:1999yj} in Bjorken-$x$, with running coupling corrections derived in Ref.~\cite{Balitsky:2006wa}. For a proton target, the initial condition for the evolution is parametrized following the McLerran-Venugopalan (MV) model~\cite{McLerran:1993ni} at $x=0.01$ as
\begin{equation}
\label{eq:dipole_ic}
    S^{(2)}_{\bm{x}_1,\bm{x}_2} = \exp \left[ -\frac{r^2 Q_{s,0}^2}{4} 
     \ln \left( \frac{1}{r \Lambda_\text{QCD}} + e \right) \right] ,
\end{equation}
with $r = |\bm{x}_1-\bm{x}_2|$, where $e$ is the Euler constant. The  parameters $Q_{s,0}^2$ and the proton transverse area $S_\perp^p$ (which enters as the normalization of the cross section) are non-perturbative inputs obtained by fitting HERA deep inelastic scattering data~\cite{Aaron:2009aa} at $x<0.01$ in Ref.~\cite{Lappi:2013zma}.
In the BK evolution the running coupling is evaluated at the scale $4C^2/(r^2\Lambda_\text{QCD}^2)$ with $\Lambda_\text{QCD}=0.241$ GeV, where $C^2$ controls the scale uncertainty in coordinate space. 

For heavier nuclei, we apply the Optical Glauber model as in \cite{Lappi:2013zma} and generalize Eq.~\eqref{eq:dipole_ic} using
\begin{equation}\label{eq:Qs0gold}
    Q_{s0}^2 \to A T_A(\bm{b}) S_\perp^p Q_{s0}^2. 
\end{equation}
Here $T_A$ is the nuclear thickness function, normalized such that $\int_{\bm{b}} T_A(\bm{b})=1$, obtained by integrating a Woods-Saxon nuclear density distribution $\rho(\bm{b}, z;R_A,a)$ along the longitudinal direction $z$. For gold the nuclear radius is $R_A=6.37$ fm, and the thickness $a=0.535$ fm. In this work we evaluate Eq.\,\eqref{eq:Qs0gold} at the average impact parameter $\langle|\bm{b}|\rangle=\int_{\bm{b}} |\bm{b}| T_A(\bm{b})$, and use the effective area $S^{\mathrm{Au}}_\perp= \pi R^2_A$.\footnote{When applied to inclusive hadron, jet, vector meson and $D$ meson production, this approach results in good agreement with LHC data~\cite{Lappi:2013zma,Ducloue:2015gfa,Ducloue:2016pqr,Ducloue:2016ywt,Mantysaari:2019nnt}.}

{\bf Cross section and elliptic anisotropy.} We present results for the angle averaged cross section and elliptic anisotropy for inclusive and diffractive dijet production in the scattering of longitudinally and transversely polarized photons with virtuality $Q^2=$ 10 GeV$^2$ off nuclear targets and center of mass energy of the photon-nucleon system $W=90$ GeV. These are defined as follows\footnote{The differential $d \Pi$ is defined as $(2 \pi)^2 |\bm{P}| \der |\bm{P}| |\bm{\Delta}| \der |\bm{\Delta}| \der y_1 \der y_2$.}:
\begin{align}
\frac{\der\sigma^{\gamma^* A \rightarrow q\bar{q} X} _{L/T}}{\der \Pi} = \int \frac{\der \theta_{\bm{P}}}{2 \pi} \frac{\der \theta_{\bm{\Delta}}}{2\pi} \frac{\der\sigma^{\gamma^* A \rightarrow q\bar{q} X} _{L/T}}{\der y_1 \der y_2 \der^2 \bm{P} \der^2 \bm{\Delta}} \,, \label{eq:avg_xsec}
\end{align} 
and
\begin{align}
    v^{\gamma^* A \rightarrow q\bar{q} X}_{2,L/T} = \frac{\int \frac{\der \theta_{\bm{P}}}{2 \pi} \frac{\der \theta_{\bm{\Delta}}}{2\pi} e^{i2(\theta_{\bm{P}}-\theta_{\bm{\Delta}})} \frac{\der\sigma^{\gamma^* A \rightarrow q\bar{q} X} _{L/T}}{\der y_1 \der y_2 \der^2 \bm{P} \der^2 \bm{\Delta}}}{ \int \frac{\der \theta_{\bm{P}}}{2 \pi} \frac{\der \theta_{\bm{\Delta}}}{2\pi} \frac{\der\sigma^{\gamma^* A \rightarrow q\bar{q} X} _{L/T}}{\der y_1 \der y_2 \der^2 \bm{P} \der^2 \bm{\Delta}}}\,. \label{eq:anisotropy}
\end{align}

We study proton and gold targets and in the inclusive case compare to the correlation limit approximation.
Additionally, we predict the ratio of diffractive to inclusive events as a function of dijet momentum for different targets and $Q^2$. All results are for fixed $z_1=z_2=0.5$.

\emph{Inclusive dijets.}
 In Fig.\,\ref{Inclusive_xsec} we present results for the angle averaged cross section Eq.\,\eqref{eq:avg_xsec} for proton (upper panels) and gold (lower panels). The panels on the left show the $|\bm{P}|$ dependence for fixed $|\bm{\Delta}|=1\,{\rm GeV}$, the contour plots (center and right) show the dependence on $|\bm{P}|$ and $|\bm{\Delta}|$ for longitudinally and transversely polarized photons.
 
 We compare the cross sections using the full multiparticle correlators Eqs.\,\eqref{Full_xsecL} and \eqref{Full_xsecT} (solid lines) and the correlation limit approximation Eqs.\,\eqref{CL_xsecL} and \eqref{CL_xsecT} in the appendix (dashed lines). The former are valid for any value of $\bm{\Delta}$, while the latter are expected to be valid only for $|\bm{P}| \gg |\bm{\Delta}|$. The expected agreement between the correlation limit and the more general result at $|\bm{P}| \gg |\bm{\Delta}|$ is clearly confirmed in all cases. Deviations from the correlation limit become large when extrapolated to the regime $|\bm{\Delta}|>|\bm{P}|$.

Importantly, we observe significant deviations from the correlation limit at $|\bm{\Delta}| < |\bm{P}| < 1.5$ GeV for the gold target, and much milder deviations for the proton. This difference is explained by saturation effects: The cross sections beyond the correlation limit approximation
receive genuine saturation corrections of order $Q^2_s / |\bm{P}|^2$ and $Q^2_s /Q^2$, in addition to kinematic corrections of order $|\bm{\Delta}|^2 / |\bm{P}|^2$ \cite{Dumitru:2016jku,Altinoluk:2019wyu}. This observation demonstrates that inclusive dijet production in e$+$A collisions at a future EIC can provide direct access to gluon saturation.
 
In Fig.\,\ref{Inclusive_v2} we present the elliptic modulation of the cross section in the angle between $\bm{P}$ and $\bm{\Delta}$ for proton (upper panels) and gold (lower panels) targets. Again, the correlation limit approximation provides a good estimate in the region $|\bm{P}|\gg|\bm{\Delta}|$, and deviations become large for $|\bm{\Delta}|\gtrsim|\bm{P}|$.
We predict a minimum $v_{2T} \sim -30\%$ for proton targets in the range $|\bm{P}|\sim|\bm{\Delta}|\sim 1.8\,{\rm GeV}$, and $v_{2T} \sim -20\%$ for gold for $|\bm{P}|\sim|\bm{\Delta}|\sim 2.2\,{\rm GeV}$. Importantly, these qualitative features are absent in the correlation limit approximation. To probe these, and the aforementioned saturation effects, experiments should focus on the kinematics $|\mathbf{P}|\sim |\mathbf{\Delta}|$.

We further confirm the large elliptic modulation for the longitudinally polarized photon, which was obtained previously in the correlation limit approximation \cite{Dumitru:2018kuw,Dumitru:2015gaa}.

\emph{Diffractive dijets.}
We show results of diffractive dijet cross sections and elliptic anisotropies for virtual photon off gold scattering in Fig.\,\ref{Diffractive_xsec}. Although our results contain only incoherent diffraction, these are the most dominant at momentum transfer $\Delta \gtrsim 1/R_A \ (\sim 0.2\,\mathrm{GeV}$ for gold), such that the result is approximately equal to the total diffractive cross section.

The cross sections exhibit different behavior depending on the polarization of the photon, in particular, the transversely polarized case shows a maximum as a function of $|\bm{P}|$, while the cross section is strictly decreasing in the longitudinal case. 

Comparing the inclusive (Fig.\,\ref{Inclusive_xsec})
and diffractive cross sections (Fig.\,\ref{Diffractive_xsec}), we observe a strong suppression of diffractive events and a different $|\bm{P}|$-dependence for the longitudinal and transverse cases. Theoretically, this can be directly related to the properties of multi-gluon correlators in the target.  
The only difference between the inclusive and diffractive cross sections are the different color structures of the correlators $\mathcal{O}$. A small dipole expansion explains the effect of this difference: The first non-vanishing term in the expansion occurs at linear order for the inclusive case and at quadratic order for diffractive production, because diffractive events require at least two gluons exchanged in the amplitude to ensure color neutrality.

The elliptic modulation of the incoherent diffractive cross section is shown in the middle panel of Fig.\,\ref{Diffractive_xsec}. For both polarizations it exhibits a sign change as a function of $|\bm{P}|$, similar to that observed in coherent diffractive dijet production \cite{Mantysaari:2019csc,Salazar:2019ncp}. 
The transverse case also shows a sign change in $|\bm{\Delta}|$ for $|\bm{P}|\gtrsim 2\,{\rm GeV}$.
Importantly, the elliptic modulation reaches large values (tens of percent) in the studied kinematic range.

In the right panels of Fig.\,\ref{Diffractive_xsec} we show the ratio of diffractive to inclusive events as a function of $|\bm{P}|$ for fixed $|\bm{\Delta}|=1.5\,{\rm GeV}$. For longitudinal polarization, the ratio is largest for $|\bm{P}|\rightarrow 0$, while there is a distinct maximum at finite $|\bm{P}|$ in the transverse case. The fraction of diffractive events increases with the target saturation scale $Q_s$ from proton to gold, and decreases with increasing photon virtuality $Q^2$. An expansion in small dipoles predicts the fraction of diffractive events to increase as $Q_{s}^2$. Using the values of $Q_{s0}^2$ from the parametrization Eq.\,\eqref{eq:Qs0gold}, we expect a factor of 2.6 increase (in the considered kinematics after BK evolution) from proton to gold. However, we find a smaller increase of 1.9 (2.3) for transversely (longitudinally) polarized photons at $|\bm{P}|\approx 1\,{\rm GeV}$ and $Q^2=4\,{\rm GeV}^2$, with a mild increase towards the expected value of 2.6 with growing $|\bm{P}|$. This behavior indicates effects of gluon saturation, which are stronger in larger nuclei. We argue that this ratio is a key measurement at a future EIC, allowing to quantify gluon saturation (differentially in $|\bm{P}|$ and $Q^2$).

{\bf Conclusions.} 
We computed inclusive and (incoherent) diffractive dijet production cross sections in e+p and e+A collisions at a future EIC within the CGC EFT.
These cross sections are sensitive probes of multi-gluon correlations inside nuclear targets at small $x$ and allow to quantitatively probe gluon saturation experimentally. 

Our approach is not restricted to the correlation limit $|\bm{P}|\gg|\bm{\Delta}|$ and significantly increases the theoretically accessible kinematic range. We employed the non-linear Gaussian approximation, using dipole correlators obtained from rcBK fits to HERA data. We validated the correlation limit approximation in inclusive dijet production for most $|\bm{P}|\gg|\bm{\Delta}|$, but found significant target dependent corrections for $|\bm{P}|\lesssim |\bm{\Delta}|$ or $|\bm{P}|\lesssim Q_s$, the latter being caused by gluon saturation effects. 
We thus argue that the regime of moderate $|\bm{P}| \sim Q_s$ of the target is particularly interesting when studying dijet production at a future EIC. Differential measurements in $\bm{P}$ and $\bm{\Delta}$ within a range that includes $Q_s$ will allow to reveal the complex multi parton structure of nuclei and uncover saturation.

We presented the first calculation of diffractive dijet cross sections and their elliptic modulation within the CGC EFT. We studied the nuclear modification of the ratio between the differential inclusive and diffractive dijet cross sections by comparing gold to proton targets at different values of $Q^2$. The dependence of the ratio between the cross sections on the target's saturation momentum indicates that saturation effects are significant in the studied kinematic regime.

In future work, we plan to include parton showers, hadronization, and full jet reconstruction. Based on results in \cite{Dumitru:2018kuw}, we expect the $v_2$ of the produced $q$-$\bar{q}$ pair presented here to be a good estimator of the observable dijet $v_2$. It will also be important to include next-to-leading order (NLO) corrections, both in small-$x$ evolution equations: NLO BK \cite{Balitsky:2008zza,Iancu:2015vea,Lappi:2015fma,Lappi:2016fmu} or NLO JIMWLK \cite{Balitsky:2013fea,Kovner:2013ona}, and the NLO impact factor \cite{Boussarie:2016bkq,Beuf:2017bpd,Hanninen:2017ddy,Roy:2019hwr,Roy:2019cux}, and to consider the effects of soft gluon radiation of the final state jets that is not captured by the jet algorithm \cite{Hatta:2019ixj}.

Detailed extraction of multi-gluon correlators in nuclei and experimental confirmation of gluon saturation will likely require complex global fits to a wide variety of experimental data. We have demonstrated that inclusive and diffractive dijet production are two of the most important processes to consider.

{\bf Acknowledgments.}
We thank Elke-Caroline Aschenauer\-, Renaud Boussarie, Kaushik Roy, S\"oren Schlichting, Vladimir Skokov, Alba Soto-Ontoso, and Raju Venugopalan for useful discussions. H.M. is supported by the Academy of Finland project 314764. N.M., F.S., and B.P.S. are supported under DOE Contract No. DE-SC0012704. N.M. is funded by the Deutsche Forschungsgemeinschaft (DFG, German Research Foundation) - project 404640738. This research used resources of the National Energy Research Scientific Computing Center, which is supported by the Office of Science of the U.S. Department of Energy under Contract No. DE-AC02-05CH11231.

\appendix
{\bf Appendix: Correlation limit approximation}
We briefly review the near back-to-back (correlation) limit approximation, $|\bm{\Delta}| \ll |\bm{P}|$. Expanding the Wilson lines to smallest order in $r$ and $r'$ one obtains \cite{Dominguez:2011wm,Dominguez:2011br}
\begin{align}
&\frac{\der \sigma^{\gamma^* A \rightarrow q\bar{q} X} _{L}}{\der y_1 \der y_2 \der ^2 \bm{P} \der ^2 \bm{\Delta}} = \alpha_{e} \alpha_s Z^2_f S_\perp   \delta_z z^3_1 z^3_2 \nonumber \\ &~~~~~~~\times \frac{8 Q^2 P^2}{(P^2 + \varepsilon_f^2)^4 } \Big( xG(|\bm{\Delta}|) +  xh(|\bm{\Delta}|) \cos 2\phi_{\bm{P}\bm{\Delta}} \Big) \,,
\label{CL_xsecL}
\end{align}
\begin{align}
&\frac{\der \sigma^{\gamma^* A \rightarrow q\bar{q} X} _{T}}{\der y_1 \der y_2 \der^2 \bm{P} \der^2 \bm{\Delta}} = \alpha_{e} \alpha_s Z^2_f S_\perp   \delta_z z_1 z_2 (z_1^2+z_2^2) \nonumber \\ &\times \frac{P^4 + \varepsilon_f^4}{(P^2 +\varepsilon_f^2)^4}  \Big(  xG(|\bm{\Delta}|) - \frac{2 P^2 \varepsilon_f^2}{P^4 + \varepsilon_f^4}  xh(|\bm{\Delta}|) \cos 2\phi_{\bm{P}\bm{\Delta}} \Big)\hspace{-0.26cm}
\label{CL_xsecT}
\end{align}
where $\phi_{\bm{P}\bm{\Delta}} = \theta_{\bm{P}}-\theta_{\bm{\Delta}}$, and $xG(|\bm{\Delta}|)$ and $xh(|\bm{\Delta}|)$ denote the trace and symmetric traceless parts of the unintegrated  Weizs\"acker-Williams gluon distribution $ xG^{jk}$ (normalized by transverse area $S_\perp$) defined as
\begin{align}
xG^{jk} = \frac{-2}{\alpha_s } \int \limits_{\bm{b}-\bm{b}'} \frac{e^{-i \bm{\Delta} \cdot (\bm{b} -\bm{b}')}}{(2\pi)^4}  \left \langle \tr[ \left( V^\dagger \partial^j V \right)_{\bm{b}} \left(V^\dagger \partial^k V \right)_{ \bm{b'}} ] \right \rangle_{x}.
\end{align}
In the Gaussian approximation \cite{Lappi:2017skr,Dumitru:2016jku}, one has
\begin{align}
\alpha_s x G = \frac{N_c^2 -1}{(2 \pi)^3 N_c} &\int_0^\infty B \der B J_0( |\bm{\Delta}|  B ) \left[ 1 - e^{-\frac{C_A}{C_F} \Gamma(B)}  \right] \nonumber \\ &~
~\times \frac{1}{\Gamma(B)} \left[ \frac{\der^2}{dB^2} + \frac{1}{B} \frac{\der}{\der B} \right] \Gamma(B)\,,
\end{align}
\begin{align}
\alpha_s x h = -\frac{N_c^2 -1}{(2 \pi)^3 N_c} &\int_0^\infty B \der B J_2( |\bm{\Delta}|  B ) \left[ 1 - e^{-\frac{C_A}{C_F} \Gamma(B)}  \right] \nonumber \\ &~~\times \frac{1}{\Gamma(B)} \left[ \frac{\der^2}{\der B^2} - \frac{1}{B} \frac{\der}{\der B} \right] \Gamma(B)
\end{align}
where $\Gamma(B) = -\ln(S^{(2)}(B))$.
Thus, the unpolarized and polarized unintegrated gluon distributions can be found from the dipole correlator $S^{(2)}$ and its derivatives.

\bibliography{ref}

\begin{thebibliography}{69}%
\makeatletter
\providecommand \@ifxundefined [1]{%
 \@ifx{#1\undefined}
}%
\providecommand \@ifnum [1]{%
 \ifnum #1\expandafter \@firstoftwo
 \else \expandafter \@secondoftwo
 \fi
}%
\providecommand \@ifx [1]{%
 \ifx #1\expandafter \@firstoftwo
 \else \expandafter \@secondoftwo
 \fi
}%
\providecommand \natexlab [1]{#1}%
\providecommand \enquote  [1]{``#1''}%
\providecommand \bibnamefont  [1]{#1}%
\providecommand \bibfnamefont [1]{#1}%
\providecommand \citenamefont [1]{#1}%
\providecommand \href@noop [0]{\@secondoftwo}%
\providecommand \href [0]{\begingroup \@sanitize@url \@href}%
\providecommand \@href[1]{\@@startlink{#1}\@@href}%
\providecommand \@@href[1]{\endgroup#1\@@endlink}%
\providecommand \@sanitize@url [0]{\catcode `\\12\catcode `\$12\catcode
  `\&12\catcode `\#12\catcode `\^12\catcode `\_12\catcode `\%12\relax}%
\providecommand \@@startlink[1]{}%
\providecommand \@@endlink[0]{}%
\providecommand \url  [0]{\begingroup\@sanitize@url \@url }%
\providecommand \@url [1]{\endgroup\@href {#1}{\urlprefix }}%
\providecommand \urlprefix  [0]{URL }%
\providecommand \Eprint [0]{\href }%
\providecommand \doibase [0]{http://dx.doi.org/}%
\providecommand \selectlanguage [0]{\@gobble}%
\providecommand \bibinfo  [0]{\@secondoftwo}%
\providecommand \bibfield  [0]{\@secondoftwo}%
\providecommand \translation [1]{[#1]}%
\providecommand \BibitemOpen [0]{}%
\providecommand \bibitemStop [0]{}%
\providecommand \bibitemNoStop [0]{.\EOS\space}%
\providecommand \EOS [0]{\spacefactor3000\relax}%
\providecommand \BibitemShut  [1]{\csname bibitem#1\endcsname}%
\let\auto@bib@innerbib\@empty
\bibitem [{\citenamefont {Boer}\ \emph {et~al.}(2011)\citenamefont {Boer} \emph
  {et~al.}}]{Boer:2011fh}%
  \BibitemOpen
  \bibfield  {author} {\bibinfo {author} {\bibfnamefont {Daniel}\ \bibnamefont
  {Boer}} \emph {et~al.},\ }\bibfield  {title} {\enquote {\bibinfo {title}
  {{Gluons and the quark sea at high energies: Distributions, polarization,
  tomography}},}\ }\href@noop {} {\  (\bibinfo {year} {2011})},\ \Eprint
  {http://arxiv.org/abs/1108.1713} {arXiv:1108.1713 [nucl-th]} \BibitemShut
  {NoStop}%
\bibitem [{\citenamefont {Accardi}\ \emph {et~al.}(2016)\citenamefont {Accardi}
  \emph {et~al.}}]{Accardi:2012qut}%
  \BibitemOpen
  \bibfield  {author} {\bibinfo {author} {\bibfnamefont {A.}~\bibnamefont
  {Accardi}} \emph {et~al.},\ }\bibfield  {title} {\enquote {\bibinfo {title}
  {{Electron Ion Collider: The Next QCD Frontier}},}\ }\href {\doibase
  10.1140/epja/i2016-16268-9} {\bibfield  {journal} {\bibinfo  {journal} {Eur.
  Phys. J.}\ }\textbf {\bibinfo {volume} {A52}},\ \bibinfo {pages} {268}
  (\bibinfo {year} {2016})},\ \Eprint {http://arxiv.org/abs/1212.1701}
  {arXiv:1212.1701 [nucl-ex]} \BibitemShut {NoStop}%
\bibitem [{\citenamefont {Aschenauer}\ \emph {et~al.}(2019)\citenamefont
  {Aschenauer}, \citenamefont {Fazio}, \citenamefont {Lee}, \citenamefont
  {Mäntysaari}, \citenamefont {Page}, \citenamefont {Schenke}, \citenamefont
  {Ullrich}, \citenamefont {Venugopalan},\ and\ \citenamefont
  {Zurita}}]{Aschenauer:2017jsk}%
  \BibitemOpen
  \bibfield  {author} {\bibinfo {author} {\bibfnamefont {E.~C.}\ \bibnamefont
  {Aschenauer}}, \bibinfo {author} {\bibfnamefont {S.}~\bibnamefont {Fazio}},
  \bibinfo {author} {\bibfnamefont {J.~H.}\ \bibnamefont {Lee}}, \bibinfo
  {author} {\bibfnamefont {H.}~\bibnamefont {Mäntysaari}}, \bibinfo {author}
  {\bibfnamefont {B.~S.}\ \bibnamefont {Page}}, \bibinfo {author}
  {\bibfnamefont {B.}~\bibnamefont {Schenke}}, \bibinfo {author} {\bibfnamefont
  {T.}~\bibnamefont {Ullrich}}, \bibinfo {author} {\bibfnamefont
  {R.}~\bibnamefont {Venugopalan}}, \ and\ \bibinfo {author} {\bibfnamefont
  {P.}~\bibnamefont {Zurita}},\ }\bibfield  {title} {\enquote {\bibinfo {title}
  {{The electron–ion collider: assessing the energy dependence of key
  measurements}},}\ }\href {\doibase 10.1088/1361-6633/aaf216} {\bibfield
  {journal} {\bibinfo  {journal} {Rept. Prog. Phys.}\ }\textbf {\bibinfo
  {volume} {82}},\ \bibinfo {pages} {024301} (\bibinfo {year} {2019})},\
  \Eprint {http://arxiv.org/abs/1708.01527} {arXiv:1708.01527 [nucl-ex]}
  \BibitemShut {NoStop}%
\bibitem [{\citenamefont {Altinoluk}\ \emph {et~al.}(2016)\citenamefont
  {Altinoluk}, \citenamefont {Armesto}, \citenamefont {Beuf},\ and\
  \citenamefont {Rezaeian}}]{Altinoluk:2015dpi}%
  \BibitemOpen
  \bibfield  {author} {\bibinfo {author} {\bibfnamefont {Tolga}\ \bibnamefont
  {Altinoluk}}, \bibinfo {author} {\bibfnamefont {Néstor}\ \bibnamefont
  {Armesto}}, \bibinfo {author} {\bibfnamefont {Guillaume}\ \bibnamefont
  {Beuf}}, \ and\ \bibinfo {author} {\bibfnamefont {Amir~H.}\ \bibnamefont
  {Rezaeian}},\ }\bibfield  {title} {\enquote {\bibinfo {title} {{Diffractive
  Dijet Production in Deep Inelastic Scattering and Photon-Hadron Collisions in
  the Color Glass Condensate}},}\ }\href {\doibase
  10.1016/j.physletb.2016.05.032} {\bibfield  {journal} {\bibinfo  {journal}
  {Phys. Lett.}\ }\textbf {\bibinfo {volume} {B758}},\ \bibinfo {pages}
  {373--383} (\bibinfo {year} {2016})},\ \Eprint
  {http://arxiv.org/abs/1511.07452} {arXiv:1511.07452 [hep-ph]} \BibitemShut
  {NoStop}%
\bibitem [{\citenamefont {Hatta}\ \emph {et~al.}(2016)\citenamefont {Hatta},
  \citenamefont {Xiao},\ and\ \citenamefont {Yuan}}]{Hatta:2016dxp}%
  \BibitemOpen
  \bibfield  {author} {\bibinfo {author} {\bibfnamefont {Yoshitaka}\
  \bibnamefont {Hatta}}, \bibinfo {author} {\bibfnamefont {Bo-Wen}\
  \bibnamefont {Xiao}}, \ and\ \bibinfo {author} {\bibfnamefont {Feng}\
  \bibnamefont {Yuan}},\ }\bibfield  {title} {\enquote {\bibinfo {title}
  {{Probing the Small-$x$ Gluon Tomography in Correlated Hard Diffractive Dijet
  Production in Deep Inelastic Scattering}},}\ }\href {\doibase
  10.1103/PhysRevLett.116.202301} {\bibfield  {journal} {\bibinfo  {journal}
  {Phys. Rev. Lett.}\ }\textbf {\bibinfo {volume} {116}},\ \bibinfo {pages}
  {202301} (\bibinfo {year} {2016})},\ \Eprint
  {http://arxiv.org/abs/1601.01585} {arXiv:1601.01585 [hep-ph]} \BibitemShut
  {NoStop}%
\bibitem [{\citenamefont {Mäntysaari}\ \emph {et~al.}(2019)\citenamefont
  {Mäntysaari}, \citenamefont {Mueller},\ and\ \citenamefont
  {Schenke}}]{Mantysaari:2019csc}%
  \BibitemOpen
  \bibfield  {author} {\bibinfo {author} {\bibfnamefont {Heikki}\ \bibnamefont
  {Mäntysaari}}, \bibinfo {author} {\bibfnamefont {Niklas}\ \bibnamefont
  {Mueller}}, \ and\ \bibinfo {author} {\bibfnamefont {Björn}\ \bibnamefont
  {Schenke}},\ }\bibfield  {title} {\enquote {\bibinfo {title} {{Diffractive
  Dijet Production and Wigner Distributions from the Color Glass
  Condensate}},}\ }\href {\doibase 10.1103/PhysRevD.99.074004} {\bibfield
  {journal} {\bibinfo  {journal} {Phys. Rev.}\ }\textbf {\bibinfo {volume}
  {D99}},\ \bibinfo {pages} {074004} (\bibinfo {year} {2019})},\ \Eprint
  {http://arxiv.org/abs/1902.05087} {arXiv:1902.05087 [hep-ph]} \BibitemShut
  {NoStop}%
\bibitem [{\citenamefont {Salazar}\ and\ \citenamefont
  {Schenke}(2019)}]{Salazar:2019ncp}%
  \BibitemOpen
  \bibfield  {author} {\bibinfo {author} {\bibfnamefont {Farid}\ \bibnamefont
  {Salazar}}\ and\ \bibinfo {author} {\bibfnamefont {Björn}\ \bibnamefont
  {Schenke}},\ }\bibfield  {title} {\enquote {\bibinfo {title} {{Diffractive
  dijet production in impact parameter dependent saturation models}},}\ }\href
  {\doibase 10.1103/PhysRevD.100.034007} {\bibfield  {journal} {\bibinfo
  {journal} {Phys. Rev.}\ }\textbf {\bibinfo {volume} {D100}},\ \bibinfo
  {pages} {034007} (\bibinfo {year} {2019})},\ \Eprint
  {http://arxiv.org/abs/1905.03763} {arXiv:1905.03763 [hep-ph]} \BibitemShut
  {NoStop}%
\bibitem [{\citenamefont {Hatta}\ \emph {et~al.}(2019)\citenamefont {Hatta},
  \citenamefont {Mueller}, \citenamefont {Ueda},\ and\ \citenamefont
  {Yuan}}]{Hatta:2019ixj}%
  \BibitemOpen
  \bibfield  {author} {\bibinfo {author} {\bibfnamefont {Yoshitaka}\
  \bibnamefont {Hatta}}, \bibinfo {author} {\bibfnamefont {Niklas}\
  \bibnamefont {Mueller}}, \bibinfo {author} {\bibfnamefont {Takahiro}\
  \bibnamefont {Ueda}}, \ and\ \bibinfo {author} {\bibfnamefont {Feng}\
  \bibnamefont {Yuan}},\ }\bibfield  {title} {\enquote {\bibinfo {title} {{QCD
  Resummation in Hard Diffractive Dijet Production at the Electron-Ion
  Collider}},}\ }\href@noop {} {\  (\bibinfo {year} {2019})},\ \Eprint
  {http://arxiv.org/abs/1907.09491} {arXiv:1907.09491 [hep-ph]} \BibitemShut
  {NoStop}%
\bibitem [{\citenamefont {Dominguez}\ \emph {et~al.}(2011)\citenamefont
  {Dominguez}, \citenamefont {Marquet}, \citenamefont {Xiao},\ and\
  \citenamefont {Yuan}}]{Dominguez:2011wm}%
  \BibitemOpen
  \bibfield  {author} {\bibinfo {author} {\bibfnamefont {Fabio}\ \bibnamefont
  {Dominguez}}, \bibinfo {author} {\bibfnamefont {Cyrille}\ \bibnamefont
  {Marquet}}, \bibinfo {author} {\bibfnamefont {Bo-Wen}\ \bibnamefont {Xiao}},
  \ and\ \bibinfo {author} {\bibfnamefont {Feng}\ \bibnamefont {Yuan}},\
  }\bibfield  {title} {\enquote {\bibinfo {title} {{Universality of
  Unintegrated Gluon Distributions at small $x$}},}\ }\href {\doibase
  10.1103/PhysRevD.83.105005} {\bibfield  {journal} {\bibinfo  {journal} {Phys.
  Rev.}\ }\textbf {\bibinfo {volume} {D83}},\ \bibinfo {pages} {105005}
  (\bibinfo {year} {2011})},\ \Eprint {http://arxiv.org/abs/1101.0715}
  {arXiv:1101.0715 [hep-ph]} \BibitemShut {NoStop}%
\bibitem [{\citenamefont {Mueller}\ \emph {et~al.}(2013)\citenamefont
  {Mueller}, \citenamefont {Xiao},\ and\ \citenamefont
  {Yuan}}]{Mueller:2013wwa}%
  \BibitemOpen
  \bibfield  {author} {\bibinfo {author} {\bibfnamefont {A.~H.}\ \bibnamefont
  {Mueller}}, \bibinfo {author} {\bibfnamefont {Bo-Wen}\ \bibnamefont {Xiao}},
  \ and\ \bibinfo {author} {\bibfnamefont {Feng}\ \bibnamefont {Yuan}},\
  }\bibfield  {title} {\enquote {\bibinfo {title} {{Sudakov double logarithms
  resummation in hard processes in the small-$x$ saturation formalism}},}\
  }\href {\doibase 10.1103/PhysRevD.88.114010} {\bibfield  {journal} {\bibinfo
  {journal} {Phys. Rev.}\ }\textbf {\bibinfo {volume} {D88}},\ \bibinfo {pages}
  {114010} (\bibinfo {year} {2013})},\ \Eprint {http://arxiv.org/abs/1308.2993}
  {arXiv:1308.2993 [hep-ph]} \BibitemShut {NoStop}%
\bibitem [{\citenamefont {Mäntysaari}\ and\ \citenamefont
  {Schenke}(2016{\natexlab{a}})}]{Mantysaari:2016ykx}%
  \BibitemOpen
  \bibfield  {author} {\bibinfo {author} {\bibfnamefont {Heikki}\ \bibnamefont
  {Mäntysaari}}\ and\ \bibinfo {author} {\bibfnamefont {Björn}\ \bibnamefont
  {Schenke}},\ }\bibfield  {title} {\enquote {\bibinfo {title} {{Evidence of
  strong proton shape fluctuations from incoherent diffraction}},}\ }\href
  {\doibase 10.1103/PhysRevLett.117.052301} {\bibfield  {journal} {\bibinfo
  {journal} {Phys. Rev. Lett.}\ }\textbf {\bibinfo {volume} {117}},\ \bibinfo
  {pages} {052301} (\bibinfo {year} {2016}{\natexlab{a}})},\ \Eprint
  {http://arxiv.org/abs/1603.04349} {arXiv:1603.04349 [hep-ph]} \BibitemShut
  {NoStop}%
\bibitem [{\citenamefont {Mäntysaari}\ and\ \citenamefont
  {Schenke}(2016{\natexlab{b}})}]{Mantysaari:2016jaz}%
  \BibitemOpen
  \bibfield  {author} {\bibinfo {author} {\bibfnamefont {Heikki}\ \bibnamefont
  {Mäntysaari}}\ and\ \bibinfo {author} {\bibfnamefont {Björn}\ \bibnamefont
  {Schenke}},\ }\bibfield  {title} {\enquote {\bibinfo {title} {{Revealing
  proton shape fluctuations with incoherent diffraction at high energy}},}\
  }\href {\doibase 10.1103/PhysRevD.94.034042} {\bibfield  {journal} {\bibinfo
  {journal} {Phys. Rev.}\ }\textbf {\bibinfo {volume} {D94}},\ \bibinfo {pages}
  {034042} (\bibinfo {year} {2016}{\natexlab{b}})},\ \Eprint
  {http://arxiv.org/abs/1607.01711} {arXiv:1607.01711 [hep-ph]} \BibitemShut
  {NoStop}%
\bibitem [{\citenamefont {Mäntysaari}\ and\ \citenamefont
  {Schenke}(2017)}]{Mantysaari:2017dwh}%
  \BibitemOpen
  \bibfield  {author} {\bibinfo {author} {\bibfnamefont {Heikki}\ \bibnamefont
  {Mäntysaari}}\ and\ \bibinfo {author} {\bibfnamefont {Björn}\ \bibnamefont
  {Schenke}},\ }\bibfield  {title} {\enquote {\bibinfo {title} {{Probing
  subnucleon scale fluctuations in ultraperipheral heavy ion collisions}},}\
  }\href {\doibase 10.1016/j.physletb.2017.07.063} {\bibfield  {journal}
  {\bibinfo  {journal} {Phys. Lett.}\ }\textbf {\bibinfo {volume} {B772}},\
  \bibinfo {pages} {832--838} (\bibinfo {year} {2017})},\ \Eprint
  {http://arxiv.org/abs/1703.09256} {arXiv:1703.09256 [hep-ph]} \BibitemShut
  {NoStop}%
\bibitem [{\citenamefont {Mäntysaari}\ and\ \citenamefont
  {Schenke}(2018)}]{Mantysaari:2018zdd}%
  \BibitemOpen
  \bibfield  {author} {\bibinfo {author} {\bibfnamefont {Heikki}\ \bibnamefont
  {Mäntysaari}}\ and\ \bibinfo {author} {\bibfnamefont {Björn}\ \bibnamefont
  {Schenke}},\ }\bibfield  {title} {\enquote {\bibinfo {title} {{Confronting
  impact parameter dependent JIMWLK evolution with HERA data}},}\ }\href
  {\doibase 10.1103/PhysRevD.98.034013} {\bibfield  {journal} {\bibinfo
  {journal} {Phys. Rev.}\ }\textbf {\bibinfo {volume} {D98}},\ \bibinfo {pages}
  {034013} (\bibinfo {year} {2018})},\ \Eprint
  {http://arxiv.org/abs/1806.06783} {arXiv:1806.06783 [hep-ph]} \BibitemShut
  {NoStop}%
\bibitem [{\citenamefont {Mäntysaari}\ and\ \citenamefont
  {Schenke}(2019)}]{Mantysaari:2019jhh}%
  \BibitemOpen
  \bibfield  {author} {\bibinfo {author} {\bibfnamefont {Heikki}\ \bibnamefont
  {Mäntysaari}}\ and\ \bibinfo {author} {\bibfnamefont {Björn}\ \bibnamefont
  {Schenke}},\ }\bibfield  {title} {\enquote {\bibinfo {title} {{Accessing the
  gluonic structure of light nuclei at the Electron Ion Collider}},}\
  }\href@noop {} {\  (\bibinfo {year} {2019})},\ \Eprint
  {http://arxiv.org/abs/1910.03297} {arXiv:1910.03297 [hep-ph]} \BibitemShut
  {NoStop}%
\bibitem [{\citenamefont {Dominguez}\ \emph {et~al.}(2012)\citenamefont
  {Dominguez}, \citenamefont {Qiu}, \citenamefont {Xiao},\ and\ \citenamefont
  {Yuan}}]{Dominguez:2011br}%
  \BibitemOpen
  \bibfield  {author} {\bibinfo {author} {\bibfnamefont {Fabio}\ \bibnamefont
  {Dominguez}}, \bibinfo {author} {\bibfnamefont {Jian-Wei}\ \bibnamefont
  {Qiu}}, \bibinfo {author} {\bibfnamefont {Bo-Wen}\ \bibnamefont {Xiao}}, \
  and\ \bibinfo {author} {\bibfnamefont {Feng}\ \bibnamefont {Yuan}},\
  }\bibfield  {title} {\enquote {\bibinfo {title} {{On the linearly polarized
  gluon distributions in the color dipole model}},}\ }\href {\doibase
  10.1103/PhysRevD.85.045003} {\bibfield  {journal} {\bibinfo  {journal} {Phys.
  Rev.}\ }\textbf {\bibinfo {volume} {D85}},\ \bibinfo {pages} {045003}
  (\bibinfo {year} {2012})},\ \Eprint {http://arxiv.org/abs/1109.6293}
  {arXiv:1109.6293 [hep-ph]} \BibitemShut {NoStop}%
\bibitem [{\citenamefont {Metz}\ and\ \citenamefont
  {Zhou}(2011)}]{Metz:2011wb}%
  \BibitemOpen
  \bibfield  {author} {\bibinfo {author} {\bibfnamefont {Andreas}\ \bibnamefont
  {Metz}}\ and\ \bibinfo {author} {\bibfnamefont {Jian}\ \bibnamefont {Zhou}},\
  }\bibfield  {title} {\enquote {\bibinfo {title} {{Distribution of linearly
  polarized gluons inside a large nucleus}},}\ }\href {\doibase
  10.1103/PhysRevD.84.051503} {\bibfield  {journal} {\bibinfo  {journal} {Phys.
  Rev.}\ }\textbf {\bibinfo {volume} {D84}},\ \bibinfo {pages} {051503}
  (\bibinfo {year} {2011})},\ \Eprint {http://arxiv.org/abs/1105.1991}
  {arXiv:1105.1991 [hep-ph]} \BibitemShut {NoStop}%
\bibitem [{\citenamefont {Dumitru}\ \emph {et~al.}(2015)\citenamefont
  {Dumitru}, \citenamefont {Lappi},\ and\ \citenamefont
  {Skokov}}]{Dumitru:2015gaa}%
  \BibitemOpen
  \bibfield  {author} {\bibinfo {author} {\bibfnamefont {Adrian}\ \bibnamefont
  {Dumitru}}, \bibinfo {author} {\bibfnamefont {Tuomas}\ \bibnamefont {Lappi}},
  \ and\ \bibinfo {author} {\bibfnamefont {Vladimir}\ \bibnamefont {Skokov}},\
  }\bibfield  {title} {\enquote {\bibinfo {title} {{Distribution of Linearly
  Polarized Gluons and Elliptic Azimuthal Anisotropy in Deep Inelastic
  Scattering Dijet Production at High Energy}},}\ }\href {\doibase
  10.1103/PhysRevLett.115.252301} {\bibfield  {journal} {\bibinfo  {journal}
  {Phys. Rev. Lett.}\ }\textbf {\bibinfo {volume} {115}},\ \bibinfo {pages}
  {252301} (\bibinfo {year} {2015})},\ \Eprint
  {http://arxiv.org/abs/1508.04438} {arXiv:1508.04438 [hep-ph]} \BibitemShut
  {NoStop}%
\bibitem [{\citenamefont {Dumitru}\ \emph {et~al.}(2019)\citenamefont
  {Dumitru}, \citenamefont {Skokov},\ and\ \citenamefont
  {Ullrich}}]{Dumitru:2018kuw}%
  \BibitemOpen
  \bibfield  {author} {\bibinfo {author} {\bibfnamefont {Adrian}\ \bibnamefont
  {Dumitru}}, \bibinfo {author} {\bibfnamefont {Vladimir}\ \bibnamefont
  {Skokov}}, \ and\ \bibinfo {author} {\bibfnamefont {Thomas}\ \bibnamefont
  {Ullrich}},\ }\bibfield  {title} {\enquote {\bibinfo {title} {{Measuring the
  Weizsäcker-Williams distribution of linearly polarized gluons at an
  electron-ion collider through dijet azimuthal asymmetries}},}\ }\href
  {\doibase 10.1103/PhysRevC.99.015204} {\bibfield  {journal} {\bibinfo
  {journal} {Phys. Rev.}\ }\textbf {\bibinfo {volume} {C99}},\ \bibinfo {pages}
  {015204} (\bibinfo {year} {2019})},\ \Eprint
  {http://arxiv.org/abs/1809.02615} {arXiv:1809.02615 [hep-ph]} \BibitemShut
  {NoStop}%
\bibitem [{\citenamefont {Dumitru}\ and\ \citenamefont
  {Skokov}(2016)}]{Dumitru:2016jku}%
  \BibitemOpen
  \bibfield  {author} {\bibinfo {author} {\bibfnamefont {Adrian}\ \bibnamefont
  {Dumitru}}\ and\ \bibinfo {author} {\bibfnamefont {Vladimir}\ \bibnamefont
  {Skokov}},\ }\bibfield  {title} {\enquote {\bibinfo {title} {{$\mathrm{cos}(4
  \phi$) azimuthal anisotropy in small-$x$ DIS dijet production beyond the
  leading power TMD limit}},}\ }\href {\doibase 10.1103/PhysRevD.94.014030}
  {\bibfield  {journal} {\bibinfo  {journal} {Phys. Rev.}\ }\textbf {\bibinfo
  {volume} {D94}},\ \bibinfo {pages} {014030} (\bibinfo {year} {2016})},\
  \Eprint {http://arxiv.org/abs/1605.02739} {arXiv:1605.02739 [hep-ph]}
  \BibitemShut {NoStop}%
\bibitem [{\citenamefont {Altinoluk}\ and\ \citenamefont
  {Boussarie}(2019)}]{Altinoluk:2019wyu}%
  \BibitemOpen
  \bibfield  {author} {\bibinfo {author} {\bibfnamefont {Tolga}\ \bibnamefont
  {Altinoluk}}\ and\ \bibinfo {author} {\bibfnamefont {Renaud}\ \bibnamefont
  {Boussarie}},\ }\bibfield  {title} {\enquote {\bibinfo {title} {{Low $x$
  physics as an infinite twist (G)TMD framework: unravelling the origins of
  saturation}},}\ }\href {\doibase 10.1007/JHEP10(2019)208} {\bibfield
  {journal} {\bibinfo  {journal} {JHEP}\ }\textbf {\bibinfo {volume} {10}},\
  \bibinfo {pages} {208} (\bibinfo {year} {2019})},\ \Eprint
  {http://arxiv.org/abs/1902.07930} {arXiv:1902.07930 [hep-ph]} \BibitemShut
  {NoStop}%
\bibitem [{\citenamefont {Altinoluk}\ \emph {et~al.}(2019)\citenamefont
  {Altinoluk}, \citenamefont {Boussarie},\ and\ \citenamefont
  {Kotko}}]{Altinoluk:2019fui}%
  \BibitemOpen
  \bibfield  {author} {\bibinfo {author} {\bibfnamefont {Tolga}\ \bibnamefont
  {Altinoluk}}, \bibinfo {author} {\bibfnamefont {Renaud}\ \bibnamefont
  {Boussarie}}, \ and\ \bibinfo {author} {\bibfnamefont {Piotr}\ \bibnamefont
  {Kotko}},\ }\bibfield  {title} {\enquote {\bibinfo {title} {{Interplay of the
  CGC and TMD frameworks to all orders in kinematic twist}},}\ }\href {\doibase
  10.1007/JHEP05(2019)156} {\bibfield  {journal} {\bibinfo  {journal} {JHEP}\
  }\textbf {\bibinfo {volume} {05}},\ \bibinfo {pages} {156} (\bibinfo {year}
  {2019})},\ \Eprint {http://arxiv.org/abs/1901.01175} {arXiv:1901.01175
  [hep-ph]} \BibitemShut {NoStop}%
\bibitem [{\citenamefont {Marquet}(2007)}]{Marquet:2007vb}%
  \BibitemOpen
  \bibfield  {author} {\bibinfo {author} {\bibfnamefont {Cyrille}\ \bibnamefont
  {Marquet}},\ }\bibfield  {title} {\enquote {\bibinfo {title} {{Forward
  inclusive dijet production and azimuthal correlations in pA collisions}},}\
  }\href {\doibase 10.1016/j.nuclphysa.2007.09.001} {\bibfield  {journal}
  {\bibinfo  {journal} {Nucl. Phys.}\ }\textbf {\bibinfo {volume} {A796}},\
  \bibinfo {pages} {41--60} (\bibinfo {year} {2007})},\ \Eprint
  {http://arxiv.org/abs/0708.0231} {arXiv:0708.0231 [hep-ph]} \BibitemShut
  {NoStop}%
\bibitem [{\citenamefont {Lappi}\ and\ \citenamefont
  {M{\"a}ntysaari}(2013{\natexlab{a}})}]{Lappi:2012nh}%
  \BibitemOpen
  \bibfield  {author} {\bibinfo {author} {\bibfnamefont {T.}~\bibnamefont
  {Lappi}}\ and\ \bibinfo {author} {\bibfnamefont {H.}~\bibnamefont
  {M{\"a}ntysaari}},\ }\bibfield  {title} {\enquote {\bibinfo {title} {{Forward
  dihadron correlations in deuteron-gold collisions with the Gaussian
  approximation of JIMWLK}},}\ }\href {\doibase
  10.1016/j.nuclphysa.2013.03.017} {\bibfield  {journal} {\bibinfo  {journal}
  {Nucl.Phys.}\ }\textbf {\bibinfo {volume} {A908}},\ \bibinfo {pages} {51--72}
  (\bibinfo {year} {2013}{\natexlab{a}})},\ \Eprint
  {http://arxiv.org/abs/1209.2853} {arXiv:1209.2853 [hep-ph]} \BibitemShut
  {NoStop}%
\bibitem [{\citenamefont {Iancu}\ and\ \citenamefont
  {Laidet}(2013)}]{Iancu:2013dta}%
  \BibitemOpen
  \bibfield  {author} {\bibinfo {author} {\bibfnamefont {Edmond}\ \bibnamefont
  {Iancu}}\ and\ \bibinfo {author} {\bibfnamefont {Julien}\ \bibnamefont
  {Laidet}},\ }\bibfield  {title} {\enquote {\bibinfo {title} {{Gluon splitting
  in a shockwave}},}\ }\href {\doibase 10.1016/j.nuclphysa.2013.07.012}
  {\bibfield  {journal} {\bibinfo  {journal} {Nucl. Phys.}\ }\textbf {\bibinfo
  {volume} {A916}},\ \bibinfo {pages} {48--78} (\bibinfo {year} {2013})},\
  \Eprint {http://arxiv.org/abs/1305.5926} {arXiv:1305.5926 [hep-ph]}
  \BibitemShut {NoStop}%
\bibitem [{\citenamefont {Kotko}\ \emph {et~al.}(2015)\citenamefont {Kotko},
  \citenamefont {Kutak}, \citenamefont {Marquet}, \citenamefont {Petreska},
  \citenamefont {Sapeta},\ and\ \citenamefont {van Hameren}}]{Kotko:2015ura}%
  \BibitemOpen
  \bibfield  {author} {\bibinfo {author} {\bibfnamefont {P.}~\bibnamefont
  {Kotko}}, \bibinfo {author} {\bibfnamefont {K.}~\bibnamefont {Kutak}},
  \bibinfo {author} {\bibfnamefont {C.}~\bibnamefont {Marquet}}, \bibinfo
  {author} {\bibfnamefont {E.}~\bibnamefont {Petreska}}, \bibinfo {author}
  {\bibfnamefont {S.}~\bibnamefont {Sapeta}}, \ and\ \bibinfo {author}
  {\bibfnamefont {A.}~\bibnamefont {van Hameren}},\ }\bibfield  {title}
  {\enquote {\bibinfo {title} {{Improved TMD factorization for forward dijet
  production in dilute-dense hadronic collisions}},}\ }\href {\doibase
  10.1007/JHEP09(2015)106} {\bibfield  {journal} {\bibinfo  {journal} {JHEP}\
  }\textbf {\bibinfo {volume} {09}},\ \bibinfo {pages} {106} (\bibinfo {year}
  {2015})},\ \Eprint {http://arxiv.org/abs/1503.03421} {arXiv:1503.03421
  [hep-ph]} \BibitemShut {NoStop}%
\bibitem [{\citenamefont {van Hameren}\ \emph {et~al.}(2016)\citenamefont {van
  Hameren}, \citenamefont {Kotko}, \citenamefont {Kutak}, \citenamefont
  {Marquet}, \citenamefont {Petreska},\ and\ \citenamefont
  {Sapeta}}]{vanHameren:2016ftb}%
  \BibitemOpen
  \bibfield  {author} {\bibinfo {author} {\bibfnamefont {A.}~\bibnamefont {van
  Hameren}}, \bibinfo {author} {\bibfnamefont {P.}~\bibnamefont {Kotko}},
  \bibinfo {author} {\bibfnamefont {K.}~\bibnamefont {Kutak}}, \bibinfo
  {author} {\bibfnamefont {C.}~\bibnamefont {Marquet}}, \bibinfo {author}
  {\bibfnamefont {E.}~\bibnamefont {Petreska}}, \ and\ \bibinfo {author}
  {\bibfnamefont {S.}~\bibnamefont {Sapeta}},\ }\bibfield  {title} {\enquote
  {\bibinfo {title} {{Forward di-jet production in p+Pb collisions in the
  small-$x$ improved TMD factorization framework}},}\ }\href {\doibase
  10.1007/JHEP12(2016)034, 10.1007/JHEP02(2019)158} {\bibfield  {journal}
  {\bibinfo  {journal} {JHEP}\ }\textbf {\bibinfo {volume} {12}},\ \bibinfo
  {pages} {034} (\bibinfo {year} {2016})},\ \bibinfo {note} {[Erratum:
  JHEP02,158(2019)]},\ \Eprint {http://arxiv.org/abs/1607.03121}
  {arXiv:1607.03121 [hep-ph]} \BibitemShut {NoStop}%
\bibitem [{\citenamefont {Albacete}\ \emph {et~al.}(2019)\citenamefont
  {Albacete}, \citenamefont {Giacalone}, \citenamefont {Marquet},\ and\
  \citenamefont {Matas}}]{Albacete:2018ruq}%
  \BibitemOpen
  \bibfield  {author} {\bibinfo {author} {\bibfnamefont {Javier~L.}\
  \bibnamefont {Albacete}}, \bibinfo {author} {\bibfnamefont {Giuliano}\
  \bibnamefont {Giacalone}}, \bibinfo {author} {\bibfnamefont {Cyrille}\
  \bibnamefont {Marquet}}, \ and\ \bibinfo {author} {\bibfnamefont {Marek}\
  \bibnamefont {Matas}},\ }\bibfield  {title} {\enquote {\bibinfo {title}
  {{Forward dihadron back-to-back correlations in $pA$ collisions}},}\ }\href
  {\doibase 10.1103/PhysRevD.99.014002} {\bibfield  {journal} {\bibinfo
  {journal} {Phys. Rev.}\ }\textbf {\bibinfo {volume} {D99}},\ \bibinfo {pages}
  {014002} (\bibinfo {year} {2019})},\ \Eprint
  {http://arxiv.org/abs/1805.05711} {arXiv:1805.05711 [hep-ph]} \BibitemShut
  {NoStop}%
\bibitem [{\citenamefont {Adare}\ \emph {et~al.}(2011)\citenamefont {Adare}
  \emph {et~al.}}]{Adare:2011sc}%
  \BibitemOpen
  \bibfield  {author} {\bibinfo {author} {\bibfnamefont {A.}~\bibnamefont
  {Adare}} \emph {et~al.} (\bibinfo {collaboration} {PHENIX}),\ }\bibfield
  {title} {\enquote {\bibinfo {title} {{Suppression of back-to-back hadron
  pairs at forward rapidity in $d+$Au Collisions at $\sqrt{s_{NN}}=200$
  GeV}},}\ }\href {\doibase 10.1103/PhysRevLett.107.172301} {\bibfield
  {journal} {\bibinfo  {journal} {Phys. Rev. Lett.}\ }\textbf {\bibinfo
  {volume} {107}},\ \bibinfo {pages} {172301} (\bibinfo {year} {2011})},\
  \Eprint {http://arxiv.org/abs/1105.5112} {arXiv:1105.5112 [nucl-ex]}
  \BibitemShut {NoStop}%
\bibitem [{\citenamefont {Braidot}(2011)}]{Braidot:2011zj}%
  \BibitemOpen
  \bibfield  {author} {\bibinfo {author} {\bibfnamefont {Ermes}\ \bibnamefont
  {Braidot}},\ }\emph {\bibinfo {title} {{Two-particle azimuthal correlations
  at forward rapidity in STAR}}},\ \href@noop {} {Ph.D. thesis},\ \bibinfo
  {school} {Utrecht U.} (\bibinfo {year} {2011}),\ \Eprint
  {http://arxiv.org/abs/1102.0931} {arXiv:1102.0931 [nucl-ex]} \BibitemShut
  {NoStop}%
\bibitem [{\citenamefont {Aktas}\ \emph {et~al.}(2007)\citenamefont {Aktas}
  \emph {et~al.}}]{Aktas:2007hn}%
  \BibitemOpen
  \bibfield  {author} {\bibinfo {author} {\bibfnamefont {A.}~\bibnamefont
  {Aktas}} \emph {et~al.} (\bibinfo {collaboration} {H1}),\ }\bibfield  {title}
  {\enquote {\bibinfo {title} {{Tests of QCD factorisation in the diffractive
  production of dijets in deep-inelastic scattering and photoproduction at
  HERA}},}\ }\href {\doibase 10.1140/epjc/s10052-007-0325-4} {\bibfield
  {journal} {\bibinfo  {journal} {Eur. Phys. J.}\ }\textbf {\bibinfo {volume}
  {C51}},\ \bibinfo {pages} {549--568} (\bibinfo {year} {2007})},\ \Eprint
  {http://arxiv.org/abs/hep-ex/0703022} {arXiv:hep-ex/0703022 [hep-ex]}
  \BibitemShut {NoStop}%
\bibitem [{\citenamefont {Guzey}\ and\ \citenamefont
  {Klasen}(2016)}]{Guzey:2016awf}%
  \BibitemOpen
  \bibfield  {author} {\bibinfo {author} {\bibfnamefont {V.}~\bibnamefont
  {Guzey}}\ and\ \bibinfo {author} {\bibfnamefont {M.}~\bibnamefont {Klasen}},\
  }\bibfield  {title} {\enquote {\bibinfo {title} {{A fresh look at
  factorization breaking in diffractive photoproduction of dijets at HERA at
  next-to-leading order QCD}},}\ }\href {\doibase
  10.1140/epjc/s10052-016-4304-5} {\bibfield  {journal} {\bibinfo  {journal}
  {Eur. Phys. J.}\ }\textbf {\bibinfo {volume} {C76}},\ \bibinfo {pages} {467}
  (\bibinfo {year} {2016})},\ \Eprint {http://arxiv.org/abs/1606.01350}
  {arXiv:1606.01350 [hep-ph]} \BibitemShut {NoStop}%
\bibitem [{\citenamefont {Helenius}\ and\ \citenamefont
  {Rasmussen}(2019)}]{Helenius:2019gbd}%
  \BibitemOpen
  \bibfield  {author} {\bibinfo {author} {\bibfnamefont {Ilkka}\ \bibnamefont
  {Helenius}}\ and\ \bibinfo {author} {\bibfnamefont {Christine~O.}\
  \bibnamefont {Rasmussen}},\ }\bibfield  {title} {\enquote {\bibinfo {title}
  {{Hard diffraction in photoproduction with Pythia 8}},}\ }\href {\doibase
  10.1140/epjc/s10052-019-6914-1} {\bibfield  {journal} {\bibinfo  {journal}
  {Eur. Phys. J.}\ }\textbf {\bibinfo {volume} {C79}},\ \bibinfo {pages} {413}
  (\bibinfo {year} {2019})},\ \Eprint {http://arxiv.org/abs/1901.05261}
  {arXiv:1901.05261 [hep-ph]} \BibitemShut {NoStop}%
\bibitem [{\citenamefont {Kowalski}\ \emph {et~al.}(2008)\citenamefont
  {Kowalski}, \citenamefont {Lappi}, \citenamefont {Marquet},\ and\
  \citenamefont {Venugopalan}}]{Kowalski:2008sa}%
  \BibitemOpen
  \bibfield  {author} {\bibinfo {author} {\bibfnamefont {H.}~\bibnamefont
  {Kowalski}}, \bibinfo {author} {\bibfnamefont {T.}~\bibnamefont {Lappi}},
  \bibinfo {author} {\bibfnamefont {C.}~\bibnamefont {Marquet}}, \ and\
  \bibinfo {author} {\bibfnamefont {R.}~\bibnamefont {Venugopalan}},\
  }\bibfield  {title} {\enquote {\bibinfo {title} {{Nuclear enhancement and
  suppression of diffractive structure functions at high energies}},}\ }\href
  {\doibase 10.1103/PhysRevC.78.045201} {\bibfield  {journal} {\bibinfo
  {journal} {Phys. Rev.}\ }\textbf {\bibinfo {volume} {C78}},\ \bibinfo {pages}
  {045201} (\bibinfo {year} {2008})},\ \Eprint {http://arxiv.org/abs/0805.4071}
  {arXiv:0805.4071 [hep-ph]} \BibitemShut {NoStop}%
\bibitem [{\citenamefont {Roy}\ and\ \citenamefont
  {Venugopalan}(2018)}]{Roy:2018jxq}%
  \BibitemOpen
  \bibfield  {author} {\bibinfo {author} {\bibfnamefont {Kaushik}\ \bibnamefont
  {Roy}}\ and\ \bibinfo {author} {\bibfnamefont {Raju}\ \bibnamefont
  {Venugopalan}},\ }\bibfield  {title} {\enquote {\bibinfo {title} {{Inclusive
  prompt photon production in electron-nucleus scattering at small $x$}},}\
  }\href {\doibase 10.1007/JHEP05(2018)013} {\bibfield  {journal} {\bibinfo
  {journal} {JHEP}\ }\textbf {\bibinfo {volume} {05}},\ \bibinfo {pages} {013}
  (\bibinfo {year} {2018})},\ \Eprint {http://arxiv.org/abs/1802.09550}
  {arXiv:1802.09550 [hep-ph]} \BibitemShut {NoStop}%
\bibitem [{\citenamefont {Lappi}\ \emph {et~al.}(2016)\citenamefont {Lappi},
  \citenamefont {Schenke}, \citenamefont {Schlichting},\ and\ \citenamefont
  {Venugopalan}}]{Lappi:2015vta}%
  \BibitemOpen
  \bibfield  {author} {\bibinfo {author} {\bibfnamefont {T.}~\bibnamefont
  {Lappi}}, \bibinfo {author} {\bibfnamefont {B.}~\bibnamefont {Schenke}},
  \bibinfo {author} {\bibfnamefont {S.}~\bibnamefont {Schlichting}}, \ and\
  \bibinfo {author} {\bibfnamefont {R.}~\bibnamefont {Venugopalan}},\
  }\bibfield  {title} {\enquote {\bibinfo {title} {{Tracing the origin of
  azimuthal gluon correlations in the color glass condensate}},}\ }\href
  {\doibase 10.1007/JHEP01(2016)061} {\bibfield  {journal} {\bibinfo  {journal}
  {JHEP}\ }\textbf {\bibinfo {volume} {01}},\ \bibinfo {pages} {061} (\bibinfo
  {year} {2016})},\ \Eprint {http://arxiv.org/abs/1509.03499} {arXiv:1509.03499
  [hep-ph]} \BibitemShut {NoStop}%
\bibitem [{\citenamefont {Dumitru}\ \emph {et~al.}(2011)\citenamefont
  {Dumitru}, \citenamefont {Jalilian-Marian}, \citenamefont {Lappi},
  \citenamefont {Schenke},\ and\ \citenamefont {Venugopalan}}]{Dumitru:2011vk}%
  \BibitemOpen
  \bibfield  {author} {\bibinfo {author} {\bibfnamefont {Adrian}\ \bibnamefont
  {Dumitru}}, \bibinfo {author} {\bibfnamefont {Jamal}\ \bibnamefont
  {Jalilian-Marian}}, \bibinfo {author} {\bibfnamefont {Tuomas}\ \bibnamefont
  {Lappi}}, \bibinfo {author} {\bibfnamefont {Bjoern}\ \bibnamefont {Schenke}},
  \ and\ \bibinfo {author} {\bibfnamefont {Raju}\ \bibnamefont {Venugopalan}},\
  }\bibfield  {title} {\enquote {\bibinfo {title} {{Renormalization group
  evolution of multi-gluon correlators in high energy QCD}},}\ }\href {\doibase
  10.1016/j.physletb.2011.11.002} {\bibfield  {journal} {\bibinfo  {journal}
  {Phys. Lett.}\ }\textbf {\bibinfo {volume} {B706}},\ \bibinfo {pages}
  {219--224} (\bibinfo {year} {2011})},\ \Eprint
  {http://arxiv.org/abs/1108.4764} {arXiv:1108.4764 [hep-ph]} \BibitemShut
  {NoStop}%
\bibitem [{\citenamefont {Jalilian-Marian}\ \emph
  {et~al.}(1997{\natexlab{a}})\citenamefont {Jalilian-Marian}, \citenamefont
  {Kovner}, \citenamefont {McLerran},\ and\ \citenamefont
  {Weigert}}]{JalilianMarian:1996xn}%
  \BibitemOpen
  \bibfield  {author} {\bibinfo {author} {\bibfnamefont {Jamal}\ \bibnamefont
  {Jalilian-Marian}}, \bibinfo {author} {\bibfnamefont {Alex}\ \bibnamefont
  {Kovner}}, \bibinfo {author} {\bibfnamefont {Larry~D.}\ \bibnamefont
  {McLerran}}, \ and\ \bibinfo {author} {\bibfnamefont {Heribert}\ \bibnamefont
  {Weigert}},\ }\bibfield  {title} {\enquote {\bibinfo {title} {{The Intrinsic
  glue distribution at very small $x$}},}\ }\href {\doibase
  10.1103/PhysRevD.55.5414} {\bibfield  {journal} {\bibinfo  {journal} {Phys.
  Rev.}\ }\textbf {\bibinfo {volume} {D55}},\ \bibinfo {pages} {5414--5428}
  (\bibinfo {year} {1997}{\natexlab{a}})},\ \Eprint
  {http://arxiv.org/abs/hep-ph/9606337} {arXiv:hep-ph/9606337 [hep-ph]}
  \BibitemShut {NoStop}%
\bibitem [{\citenamefont {Jalilian-Marian}\ \emph {et~al.}(1998)\citenamefont
  {Jalilian-Marian}, \citenamefont {Kovner}, \citenamefont {Leonidov},\ and\
  \citenamefont {Weigert}}]{JalilianMarian:1997gr}%
  \BibitemOpen
  \bibfield  {author} {\bibinfo {author} {\bibfnamefont {Jamal}\ \bibnamefont
  {Jalilian-Marian}}, \bibinfo {author} {\bibfnamefont {Alex}\ \bibnamefont
  {Kovner}}, \bibinfo {author} {\bibfnamefont {Andrei}\ \bibnamefont
  {Leonidov}}, \ and\ \bibinfo {author} {\bibfnamefont {Heribert}\ \bibnamefont
  {Weigert}},\ }\bibfield  {title} {\enquote {\bibinfo {title} {{The Wilson
  renormalization group for low $x$ physics: Towards the high density
  regime}},}\ }\href {\doibase 10.1103/PhysRevD.59.014014} {\bibfield
  {journal} {\bibinfo  {journal} {Phys. Rev.}\ }\textbf {\bibinfo {volume}
  {D59}},\ \bibinfo {pages} {014014} (\bibinfo {year} {1998})},\ \Eprint
  {http://arxiv.org/abs/hep-ph/9706377} {arXiv:hep-ph/9706377 [hep-ph]}
  \BibitemShut {NoStop}%
\bibitem [{\citenamefont {Jalilian-Marian}\ \emph
  {et~al.}(1997{\natexlab{b}})\citenamefont {Jalilian-Marian}, \citenamefont
  {Kovner}, \citenamefont {Leonidov},\ and\ \citenamefont
  {Weigert}}]{JalilianMarian:1997jx}%
  \BibitemOpen
  \bibfield  {author} {\bibinfo {author} {\bibfnamefont {Jamal}\ \bibnamefont
  {Jalilian-Marian}}, \bibinfo {author} {\bibfnamefont {Alex}\ \bibnamefont
  {Kovner}}, \bibinfo {author} {\bibfnamefont {Andrei}\ \bibnamefont
  {Leonidov}}, \ and\ \bibinfo {author} {\bibfnamefont {Heribert}\ \bibnamefont
  {Weigert}},\ }\bibfield  {title} {\enquote {\bibinfo {title} {{The BFKL
  equation from the Wilson renormalization group}},}\ }\href {\doibase
  10.1016/S0550-3213(97)00440-9} {\bibfield  {journal} {\bibinfo  {journal}
  {Nucl. Phys.}\ }\textbf {\bibinfo {volume} {B504}},\ \bibinfo {pages}
  {415--431} (\bibinfo {year} {1997}{\natexlab{b}})},\ \Eprint
  {http://arxiv.org/abs/hep-ph/9701284} {arXiv:hep-ph/9701284 [hep-ph]}
  \BibitemShut {NoStop}%
\bibitem [{\citenamefont {Iancu}\ \emph
  {et~al.}(2001{\natexlab{a}})\citenamefont {Iancu}, \citenamefont {Leonidov},\
  and\ \citenamefont {McLerran}}]{Iancu:2001ad}%
  \BibitemOpen
  \bibfield  {author} {\bibinfo {author} {\bibfnamefont {Edmond}\ \bibnamefont
  {Iancu}}, \bibinfo {author} {\bibfnamefont {Andrei}\ \bibnamefont
  {Leonidov}}, \ and\ \bibinfo {author} {\bibfnamefont {Larry~D.}\ \bibnamefont
  {McLerran}},\ }\bibfield  {title} {\enquote {\bibinfo {title} {{The
  Renormalization group equation for the color glass condensate}},}\ }\href
  {\doibase 10.1016/S0370-2693(01)00524-X} {\bibfield  {journal} {\bibinfo
  {journal} {Phys. Lett.}\ }\textbf {\bibinfo {volume} {B510}},\ \bibinfo
  {pages} {133--144} (\bibinfo {year} {2001}{\natexlab{a}})},\ \Eprint
  {http://arxiv.org/abs/hep-ph/0102009} {arXiv:hep-ph/0102009 [hep-ph]}
  \BibitemShut {NoStop}%
\bibitem [{\citenamefont {Iancu}\ \emph
  {et~al.}(2001{\natexlab{b}})\citenamefont {Iancu}, \citenamefont {Leonidov},\
  and\ \citenamefont {McLerran}}]{Iancu:2000hn}%
  \BibitemOpen
  \bibfield  {author} {\bibinfo {author} {\bibfnamefont {Edmond}\ \bibnamefont
  {Iancu}}, \bibinfo {author} {\bibfnamefont {Andrei}\ \bibnamefont
  {Leonidov}}, \ and\ \bibinfo {author} {\bibfnamefont {Larry~D.}\ \bibnamefont
  {McLerran}},\ }\bibfield  {title} {\enquote {\bibinfo {title} {{Nonlinear
  gluon evolution in the color glass condensate. 1.}}}\ }\href {\doibase
  10.1016/S0375-9474(01)00642-X} {\bibfield  {journal} {\bibinfo  {journal}
  {Nucl. Phys.}\ }\textbf {\bibinfo {volume} {A692}},\ \bibinfo {pages}
  {583--645} (\bibinfo {year} {2001}{\natexlab{b}})},\ \Eprint
  {http://arxiv.org/abs/hep-ph/0011241} {arXiv:hep-ph/0011241 [hep-ph]}
  \BibitemShut {NoStop}%
\bibitem [{\citenamefont {Ferreiro}\ \emph {et~al.}(2002)\citenamefont
  {Ferreiro}, \citenamefont {Iancu}, \citenamefont {Leonidov},\ and\
  \citenamefont {McLerran}}]{Ferreiro:2001qy}%
  \BibitemOpen
  \bibfield  {author} {\bibinfo {author} {\bibfnamefont {Elena}\ \bibnamefont
  {Ferreiro}}, \bibinfo {author} {\bibfnamefont {Edmond}\ \bibnamefont
  {Iancu}}, \bibinfo {author} {\bibfnamefont {Andrei}\ \bibnamefont
  {Leonidov}}, \ and\ \bibinfo {author} {\bibfnamefont {Larry}\ \bibnamefont
  {McLerran}},\ }\bibfield  {title} {\enquote {\bibinfo {title} {{Nonlinear
  gluon evolution in the color glass condensate. 2.}}}\ }\href {\doibase
  10.1016/S0375-9474(01)01329-X} {\bibfield  {journal} {\bibinfo  {journal}
  {Nucl. Phys.}\ }\textbf {\bibinfo {volume} {A703}},\ \bibinfo {pages}
  {489--538} (\bibinfo {year} {2002})},\ \Eprint
  {http://arxiv.org/abs/hep-ph/0109115} {arXiv:hep-ph/0109115 [hep-ph]}
  \BibitemShut {NoStop}%
\bibitem [{\citenamefont {Iancu}\ and\ \citenamefont
  {McLerran}(2001)}]{Iancu:2001md}%
  \BibitemOpen
  \bibfield  {author} {\bibinfo {author} {\bibfnamefont {Edmond}\ \bibnamefont
  {Iancu}}\ and\ \bibinfo {author} {\bibfnamefont {Larry~D.}\ \bibnamefont
  {McLerran}},\ }\bibfield  {title} {\enquote {\bibinfo {title} {{Saturation
  and universality in QCD at small $x$}},}\ }\href {\doibase
  10.1016/S0370-2693(01)00526-3} {\bibfield  {journal} {\bibinfo  {journal}
  {Phys. Lett.}\ }\textbf {\bibinfo {volume} {B510}},\ \bibinfo {pages}
  {145--154} (\bibinfo {year} {2001})},\ \Eprint
  {http://arxiv.org/abs/hep-ph/0103032} {arXiv:hep-ph/0103032 [hep-ph]}
  \BibitemShut {NoStop}%
\bibitem [{\citenamefont {Dominguez}\ \emph {et~al.}(2009)\citenamefont
  {Dominguez}, \citenamefont {Marquet},\ and\ \citenamefont
  {Wu}}]{Dominguez:2008aa}%
  \BibitemOpen
  \bibfield  {author} {\bibinfo {author} {\bibfnamefont {Fabio}\ \bibnamefont
  {Dominguez}}, \bibinfo {author} {\bibfnamefont {Cyrille}\ \bibnamefont
  {Marquet}}, \ and\ \bibinfo {author} {\bibfnamefont {Bin}\ \bibnamefont
  {Wu}},\ }\bibfield  {title} {\enquote {\bibinfo {title} {{On multiple
  scatterings of mesons in hot and cold QCD matter}},}\ }\href {\doibase
  10.1016/j.nuclphysa.2009.03.008} {\bibfield  {journal} {\bibinfo  {journal}
  {Nucl. Phys.}\ }\textbf {\bibinfo {volume} {A823}},\ \bibinfo {pages}
  {99--119} (\bibinfo {year} {2009})},\ \Eprint
  {http://arxiv.org/abs/0812.3878} {arXiv:0812.3878 [nucl-th]} \BibitemShut
  {NoStop}%
\bibitem [{\citenamefont {Blaizot}\ \emph {et~al.}(2004)\citenamefont
  {Blaizot}, \citenamefont {Gelis},\ and\ \citenamefont
  {Venugopalan}}]{Blaizot:2004wv}%
  \BibitemOpen
  \bibfield  {author} {\bibinfo {author} {\bibfnamefont {Jean~Paul}\
  \bibnamefont {Blaizot}}, \bibinfo {author} {\bibfnamefont {Francois}\
  \bibnamefont {Gelis}}, \ and\ \bibinfo {author} {\bibfnamefont {Raju}\
  \bibnamefont {Venugopalan}},\ }\bibfield  {title} {\enquote {\bibinfo {title}
  {{High-energy pA collisions in the color glass condensate approach. 2. Quark
  production}},}\ }\href {\doibase 10.1016/j.nuclphysa.2004.07.006} {\bibfield
  {journal} {\bibinfo  {journal} {Nucl. Phys.}\ }\textbf {\bibinfo {volume}
  {A743}},\ \bibinfo {pages} {57--91} (\bibinfo {year} {2004})},\ \Eprint
  {http://arxiv.org/abs/hep-ph/0402257} {arXiv:hep-ph/0402257 [hep-ph]}
  \BibitemShut {NoStop}%
\bibitem [{\citenamefont {Fukushima}\ and\ \citenamefont
  {Hidaka}(2007)}]{Fukushima:2007dy}%
  \BibitemOpen
  \bibfield  {author} {\bibinfo {author} {\bibfnamefont {Kenji}\ \bibnamefont
  {Fukushima}}\ and\ \bibinfo {author} {\bibfnamefont {Yoshimasa}\ \bibnamefont
  {Hidaka}},\ }\bibfield  {title} {\enquote {\bibinfo {title} {{Light
  projectile scattering off the color glass condensate}},}\ }\href {\doibase
  10.1088/1126-6708/2007/06/040} {\bibfield  {journal} {\bibinfo  {journal}
  {JHEP}\ }\textbf {\bibinfo {volume} {06}},\ \bibinfo {pages} {040} (\bibinfo
  {year} {2007})},\ \Eprint {http://arxiv.org/abs/0704.2806} {arXiv:0704.2806
  [hep-ph]} \BibitemShut {NoStop}%
\bibitem [{\citenamefont {Balitsky}(1996)}]{Balitsky:1995ub}%
  \BibitemOpen
  \bibfield  {author} {\bibinfo {author} {\bibfnamefont {I.}~\bibnamefont
  {Balitsky}},\ }\bibfield  {title} {\enquote {\bibinfo {title} {{Operator
  expansion for high-energy scattering}},}\ }\href {\doibase
  10.1016/0550-3213(95)00638-9} {\bibfield  {journal} {\bibinfo  {journal}
  {Nucl. Phys.}\ }\textbf {\bibinfo {volume} {B463}},\ \bibinfo {pages}
  {99--160} (\bibinfo {year} {1996})},\ \Eprint
  {http://arxiv.org/abs/hep-ph/9509348} {arXiv:hep-ph/9509348 [hep-ph]}
  \BibitemShut {NoStop}%
\bibitem [{\citenamefont {Kovchegov}(1999)}]{Kovchegov:1999yj}%
  \BibitemOpen
  \bibfield  {author} {\bibinfo {author} {\bibfnamefont {Yuri~V.}\ \bibnamefont
  {Kovchegov}},\ }\bibfield  {title} {\enquote {\bibinfo {title} {{Small $x$
  $F_2$ structure function of a nucleus including multiple pomeron
  exchanges}},}\ }\href {\doibase 10.1103/PhysRevD.60.034008} {\bibfield
  {journal} {\bibinfo  {journal} {Phys. Rev.}\ }\textbf {\bibinfo {volume}
  {D60}},\ \bibinfo {pages} {034008} (\bibinfo {year} {1999})},\ \Eprint
  {http://arxiv.org/abs/hep-ph/9901281} {arXiv:hep-ph/9901281 [hep-ph]}
  \BibitemShut {NoStop}%
\bibitem [{\citenamefont {Balitsky}(2007)}]{Balitsky:2006wa}%
  \BibitemOpen
  \bibfield  {author} {\bibinfo {author} {\bibfnamefont {Ian}\ \bibnamefont
  {Balitsky}},\ }\bibfield  {title} {\enquote {\bibinfo {title} {{Quark
  contribution to the small-$x$ evolution of color dipole}},}\ }\href {\doibase
  10.1103/PhysRevD.75.014001} {\bibfield  {journal} {\bibinfo  {journal} {Phys.
  Rev.}\ }\textbf {\bibinfo {volume} {D75}},\ \bibinfo {pages} {014001}
  (\bibinfo {year} {2007})},\ \Eprint {http://arxiv.org/abs/hep-ph/0609105}
  {arXiv:hep-ph/0609105 [hep-ph]} \BibitemShut {NoStop}%
\bibitem [{\citenamefont {McLerran}\ and\ \citenamefont
  {Venugopalan}(1994)}]{McLerran:1993ni}%
  \BibitemOpen
  \bibfield  {author} {\bibinfo {author} {\bibfnamefont {Larry~D.}\
  \bibnamefont {McLerran}}\ and\ \bibinfo {author} {\bibfnamefont {Raju}\
  \bibnamefont {Venugopalan}},\ }\bibfield  {title} {\enquote {\bibinfo {title}
  {{Computing quark and gluon distribution functions for very large nuclei}},}\
  }\href {\doibase 10.1103/PhysRevD.49.2233} {\bibfield  {journal} {\bibinfo
  {journal} {Phys. Rev.}\ }\textbf {\bibinfo {volume} {D49}},\ \bibinfo {pages}
  {2233--2241} (\bibinfo {year} {1994})},\ \Eprint
  {http://arxiv.org/abs/hep-ph/9309289} {arXiv:hep-ph/9309289 [hep-ph]}
  \BibitemShut {NoStop}%
\bibitem [{\citenamefont {Aaron}\ \emph {et~al.}(2010)\citenamefont {Aaron}
  \emph {et~al.}}]{Aaron:2009aa}%
  \BibitemOpen
  \bibfield  {author} {\bibinfo {author} {\bibfnamefont {F.D.}\ \bibnamefont
  {Aaron}} \emph {et~al.} (\bibinfo {collaboration} {H1 and ZEUS}),\ }\bibfield
   {title} {\enquote {\bibinfo {title} {{Combined Measurement and QCD Analysis
  of the Inclusive $e^\pm p$ Scattering Cross Sections at HERA}},}\ }\href
  {\doibase 10.1007/JHEP01(2010)109} {\bibfield  {journal} {\bibinfo  {journal}
  {JHEP}\ }\textbf {\bibinfo {volume} {1001}},\ \bibinfo {pages} {109}
  (\bibinfo {year} {2010})},\ \Eprint {http://arxiv.org/abs/0911.0884}
  {arXiv:0911.0884 [hep-ex]} \BibitemShut {NoStop}%
\bibitem [{\citenamefont {Lappi}\ and\ \citenamefont
  {M{\"a}ntysaari}(2013{\natexlab{b}})}]{Lappi:2013zma}%
  \BibitemOpen
  \bibfield  {author} {\bibinfo {author} {\bibfnamefont {T.}~\bibnamefont
  {Lappi}}\ and\ \bibinfo {author} {\bibfnamefont {H.}~\bibnamefont
  {M{\"a}ntysaari}},\ }\bibfield  {title} {\enquote {\bibinfo {title} {{Single
  inclusive particle production at high energy from HERA data to proton-nucleus
  collisions}},}\ }\href {\doibase 10.1103/PhysRevD.88.114020} {\bibfield
  {journal} {\bibinfo  {journal} {Phys. Rev.}\ }\textbf {\bibinfo {volume}
  {D88}},\ \bibinfo {pages} {114020} (\bibinfo {year} {2013}{\natexlab{b}})},\
  \Eprint {http://arxiv.org/abs/1309.6963} {arXiv:1309.6963 [hep-ph]}
  \BibitemShut {NoStop}%
\bibitem [{\citenamefont {Ducloué}\ \emph {et~al.}(2015)\citenamefont
  {Ducloué}, \citenamefont {Lappi},\ and\ \citenamefont
  {Mäntysaari}}]{Ducloue:2015gfa}%
  \BibitemOpen
  \bibfield  {author} {\bibinfo {author} {\bibfnamefont {B.}~\bibnamefont
  {Ducloué}}, \bibinfo {author} {\bibfnamefont {T.}~\bibnamefont {Lappi}}, \
  and\ \bibinfo {author} {\bibfnamefont {H.}~\bibnamefont {Mäntysaari}},\
  }\bibfield  {title} {\enquote {\bibinfo {title} {{Forward $J/\psi$ production
  in proton-nucleus collisions at high energy}},}\ }\href {\doibase
  10.1103/PhysRevD.91.114005} {\bibfield  {journal} {\bibinfo  {journal} {Phys.
  Rev.}\ }\textbf {\bibinfo {volume} {D91}},\ \bibinfo {pages} {114005}
  (\bibinfo {year} {2015})},\ \Eprint {http://arxiv.org/abs/1503.02789}
  {arXiv:1503.02789 [hep-ph]} \BibitemShut {NoStop}%
\bibitem [{\citenamefont {Ducloué}\ \emph {et~al.}(2016)\citenamefont
  {Ducloué}, \citenamefont {Lappi},\ and\ \citenamefont
  {Mäntysaari}}]{Ducloue:2016pqr}%
  \BibitemOpen
  \bibfield  {author} {\bibinfo {author} {\bibfnamefont {B.}~\bibnamefont
  {Ducloué}}, \bibinfo {author} {\bibfnamefont {T.}~\bibnamefont {Lappi}}, \
  and\ \bibinfo {author} {\bibfnamefont {H.}~\bibnamefont {Mäntysaari}},\
  }\bibfield  {title} {\enquote {\bibinfo {title} {{Forward $J/\psi$ production
  at high energy: centrality dependence and mean transverse momentum}},}\
  }\href {\doibase 10.1103/PhysRevD.94.074031} {\bibfield  {journal} {\bibinfo
  {journal} {Phys. Rev.}\ }\textbf {\bibinfo {volume} {D94}},\ \bibinfo {pages}
  {074031} (\bibinfo {year} {2016})},\ \Eprint
  {http://arxiv.org/abs/1605.05680} {arXiv:1605.05680 [hep-ph]} \BibitemShut
  {NoStop}%
\bibitem [{\citenamefont {Ducloué}\ \emph {et~al.}(2017)\citenamefont
  {Ducloué}, \citenamefont {Lappi},\ and\ \citenamefont
  {Mäntysaari}}]{Ducloue:2016ywt}%
  \BibitemOpen
  \bibfield  {author} {\bibinfo {author} {\bibfnamefont {B.}~\bibnamefont
  {Ducloué}}, \bibinfo {author} {\bibfnamefont {T.}~\bibnamefont {Lappi}}, \
  and\ \bibinfo {author} {\bibfnamefont {H.}~\bibnamefont {Mäntysaari}},\
  }\bibfield  {title} {\enquote {\bibinfo {title} {{Forward $J/\psi$ and $D$
  meson nuclear suppression at the LHC}},}\ }\href {\doibase
  10.1016/j.nuclphysbps.2017.05.071} {\bibfield  {journal} {\bibinfo  {journal}
  {Nucl. Part. Phys. Proc.}\ }\textbf {\bibinfo {volume} {289-290}},\ \bibinfo
  {pages} {309--312} (\bibinfo {year} {2017})},\ \Eprint
  {http://arxiv.org/abs/1612.04585} {arXiv:1612.04585 [hep-ph]} \BibitemShut
  {NoStop}%
\bibitem [{\citenamefont {Mäntysaari}\ and\ \citenamefont
  {Paukkunen}(2019)}]{Mantysaari:2019nnt}%
  \BibitemOpen
  \bibfield  {author} {\bibinfo {author} {\bibfnamefont {Heikki}\ \bibnamefont
  {Mäntysaari}}\ and\ \bibinfo {author} {\bibfnamefont {Hannu}\ \bibnamefont
  {Paukkunen}},\ }\bibfield  {title} {\enquote {\bibinfo {title} {{Saturation
  and forward jets in proton-lead collisions at the LHC}},}\ }\href@noop {} {\
  (\bibinfo {year} {2019})},\ \Eprint {http://arxiv.org/abs/1910.13116}
  {arXiv:1910.13116 [hep-ph]} \BibitemShut {NoStop}%
\bibitem [{\citenamefont {Balitsky}\ and\ \citenamefont
  {Chirilli}(2008)}]{Balitsky:2008zza}%
  \BibitemOpen
  \bibfield  {author} {\bibinfo {author} {\bibfnamefont {Ian}\ \bibnamefont
  {Balitsky}}\ and\ \bibinfo {author} {\bibfnamefont {Giovanni~A.}\
  \bibnamefont {Chirilli}},\ }\bibfield  {title} {\enquote {\bibinfo {title}
  {{Next-to-leading order evolution of color dipoles}},}\ }\href {\doibase
  10.1103/PhysRevD.77.014019} {\bibfield  {journal} {\bibinfo  {journal} {Phys.
  Rev.}\ }\textbf {\bibinfo {volume} {D77}},\ \bibinfo {pages} {014019}
  (\bibinfo {year} {2008})},\ \Eprint {http://arxiv.org/abs/0710.4330}
  {arXiv:0710.4330 [hep-ph]} \BibitemShut {NoStop}%
\bibitem [{\citenamefont {Iancu}\ \emph {et~al.}(2015)\citenamefont {Iancu},
  \citenamefont {Madrigal}, \citenamefont {Mueller}, \citenamefont {Soyez},\
  and\ \citenamefont {Triantafyllopoulos}}]{Iancu:2015vea}%
  \BibitemOpen
  \bibfield  {author} {\bibinfo {author} {\bibfnamefont {E.}~\bibnamefont
  {Iancu}}, \bibinfo {author} {\bibfnamefont {J.D.}\ \bibnamefont {Madrigal}},
  \bibinfo {author} {\bibfnamefont {A.H.}\ \bibnamefont {Mueller}}, \bibinfo
  {author} {\bibfnamefont {G.}~\bibnamefont {Soyez}}, \ and\ \bibinfo {author}
  {\bibfnamefont {D.N.}\ \bibnamefont {Triantafyllopoulos}},\ }\bibfield
  {title} {\enquote {\bibinfo {title} {{Resumming double logarithms in the QCD
  evolution of color dipoles}},}\ }\href {\doibase
  10.1016/j.physletb.2015.03.068} {\bibfield  {journal} {\bibinfo  {journal}
  {Phys. Lett.}\ }\textbf {\bibinfo {volume} {B744}},\ \bibinfo {pages}
  {293--302} (\bibinfo {year} {2015})},\ \Eprint
  {http://arxiv.org/abs/1502.05642} {arXiv:1502.05642 [hep-ph]} \BibitemShut
  {NoStop}%
\bibitem [{\citenamefont {Lappi}\ and\ \citenamefont
  {Mäntysaari}(2015)}]{Lappi:2015fma}%
  \BibitemOpen
  \bibfield  {author} {\bibinfo {author} {\bibfnamefont {T.}~\bibnamefont
  {Lappi}}\ and\ \bibinfo {author} {\bibfnamefont {H.}~\bibnamefont
  {Mäntysaari}},\ }\bibfield  {title} {\enquote {\bibinfo {title} {{Direct
  numerical solution of the coordinate space Balitsky-Kovchegov equation at
  next to leading order}},}\ }\href {\doibase 10.1103/PhysRevD.91.074016}
  {\bibfield  {journal} {\bibinfo  {journal} {Phys. Rev.}\ }\textbf {\bibinfo
  {volume} {D91}},\ \bibinfo {pages} {074016} (\bibinfo {year} {2015})},\
  \Eprint {http://arxiv.org/abs/1502.02400} {arXiv:1502.02400 [hep-ph]}
  \BibitemShut {NoStop}%
\bibitem [{\citenamefont {Lappi}\ and\ \citenamefont
  {Mäntysaari}(2016)}]{Lappi:2016fmu}%
  \BibitemOpen
  \bibfield  {author} {\bibinfo {author} {\bibfnamefont {T.}~\bibnamefont
  {Lappi}}\ and\ \bibinfo {author} {\bibfnamefont {H.}~\bibnamefont
  {Mäntysaari}},\ }\bibfield  {title} {\enquote {\bibinfo {title}
  {{Next-to-leading order Balitsky-Kovchegov equation with resummation}},}\
  }\href {\doibase 10.1103/PhysRevD.93.094004} {\bibfield  {journal} {\bibinfo
  {journal} {Phys. Rev.}\ }\textbf {\bibinfo {volume} {D93}},\ \bibinfo {pages}
  {094004} (\bibinfo {year} {2016})},\ \Eprint
  {http://arxiv.org/abs/1601.06598} {arXiv:1601.06598 [hep-ph]} \BibitemShut
  {NoStop}%
\bibitem [{\citenamefont {Balitsky}\ and\ \citenamefont
  {Chirilli}(2013)}]{Balitsky:2013fea}%
  \BibitemOpen
  \bibfield  {author} {\bibinfo {author} {\bibfnamefont {Ian}\ \bibnamefont
  {Balitsky}}\ and\ \bibinfo {author} {\bibfnamefont {Giovanni~A.}\
  \bibnamefont {Chirilli}},\ }\bibfield  {title} {\enquote {\bibinfo {title}
  {{Rapidity evolution of Wilson lines at the next-to-leading order}},}\ }\href
  {\doibase 10.1103/PhysRevD.88.111501} {\bibfield  {journal} {\bibinfo
  {journal} {Phys. Rev.}\ }\textbf {\bibinfo {volume} {D88}},\ \bibinfo {pages}
  {111501} (\bibinfo {year} {2013})},\ \Eprint {http://arxiv.org/abs/1309.7644}
  {arXiv:1309.7644 [hep-ph]} \BibitemShut {NoStop}%
\bibitem [{\citenamefont {Kovner}\ \emph {et~al.}(2014)\citenamefont {Kovner},
  \citenamefont {Lublinsky},\ and\ \citenamefont {Mulian}}]{Kovner:2013ona}%
  \BibitemOpen
  \bibfield  {author} {\bibinfo {author} {\bibfnamefont {Alex}\ \bibnamefont
  {Kovner}}, \bibinfo {author} {\bibfnamefont {Michael}\ \bibnamefont
  {Lublinsky}}, \ and\ \bibinfo {author} {\bibfnamefont {Yair}\ \bibnamefont
  {Mulian}},\ }\bibfield  {title} {\enquote {\bibinfo {title}
  {{Jalilian-Marian, Iancu, McLerran, Weigert, Leonidov, Kovner evolution at
  next to leading order}},}\ }\href {\doibase 10.1103/PhysRevD.89.061704}
  {\bibfield  {journal} {\bibinfo  {journal} {Phys. Rev.}\ }\textbf {\bibinfo
  {volume} {D89}},\ \bibinfo {pages} {061704} (\bibinfo {year} {2014})},\
  \Eprint {http://arxiv.org/abs/1310.0378} {arXiv:1310.0378 [hep-ph]}
  \BibitemShut {NoStop}%
\bibitem [{\citenamefont {Boussarie}\ \emph {et~al.}(2017)\citenamefont
  {Boussarie}, \citenamefont {Grabovsky}, \citenamefont {Ivanov}, \citenamefont
  {Szymanowski},\ and\ \citenamefont {Wallon}}]{Boussarie:2016bkq}%
  \BibitemOpen
  \bibfield  {author} {\bibinfo {author} {\bibfnamefont {R.}~\bibnamefont
  {Boussarie}}, \bibinfo {author} {\bibfnamefont {A.~V.}\ \bibnamefont
  {Grabovsky}}, \bibinfo {author} {\bibfnamefont {D.~{\relax Yu}.}\
  \bibnamefont {Ivanov}}, \bibinfo {author} {\bibfnamefont {L.}~\bibnamefont
  {Szymanowski}}, \ and\ \bibinfo {author} {\bibfnamefont {S.}~\bibnamefont
  {Wallon}},\ }\bibfield  {title} {\enquote {\bibinfo {title} {{Next-to-Leading
  Order Computation of Exclusive Diffractive Light Vector Meson Production in a
  Saturation Framework}},}\ }\href {\doibase 10.1103/PhysRevLett.119.072002}
  {\bibfield  {journal} {\bibinfo  {journal} {Phys. Rev. Lett.}\ }\textbf
  {\bibinfo {volume} {119}},\ \bibinfo {pages} {072002} (\bibinfo {year}
  {2017})},\ \Eprint {http://arxiv.org/abs/1612.08026} {arXiv:1612.08026
  [hep-ph]} \BibitemShut {NoStop}%
\bibitem [{\citenamefont {Beuf}(2017)}]{Beuf:2017bpd}%
  \BibitemOpen
  \bibfield  {author} {\bibinfo {author} {\bibfnamefont {Guillaume}\
  \bibnamefont {Beuf}},\ }\bibfield  {title} {\enquote {\bibinfo {title}
  {{Dipole factorization for DIS at NLO: Combining the $q\bar{q}$ and
  $q\bar{q}g$ contributions}},}\ }\href {\doibase 10.1103/PhysRevD.96.074033}
  {\bibfield  {journal} {\bibinfo  {journal} {Phys. Rev.}\ }\textbf {\bibinfo
  {volume} {D96}},\ \bibinfo {pages} {074033} (\bibinfo {year} {2017})},\
  \Eprint {http://arxiv.org/abs/1708.06557} {arXiv:1708.06557 [hep-ph]}
  \BibitemShut {NoStop}%
\bibitem [{\citenamefont {Hänninen}\ \emph {et~al.}(2018)\citenamefont
  {Hänninen}, \citenamefont {Lappi},\ and\ \citenamefont
  {Paatelainen}}]{Hanninen:2017ddy}%
  \BibitemOpen
  \bibfield  {author} {\bibinfo {author} {\bibfnamefont {H.}~\bibnamefont
  {Hänninen}}, \bibinfo {author} {\bibfnamefont {T.}~\bibnamefont {Lappi}}, \
  and\ \bibinfo {author} {\bibfnamefont {R.}~\bibnamefont {Paatelainen}},\
  }\bibfield  {title} {\enquote {\bibinfo {title} {{One-loop corrections to
  light cone wave functions: the dipole picture DIS cross section}},}\ }\href
  {\doibase 10.1016/j.aop.2018.04.015} {\bibfield  {journal} {\bibinfo
  {journal} {Annals Phys.}\ }\textbf {\bibinfo {volume} {393}},\ \bibinfo
  {pages} {358--412} (\bibinfo {year} {2018})},\ \Eprint
  {http://arxiv.org/abs/1711.08207} {arXiv:1711.08207 [hep-ph]} \BibitemShut
  {NoStop}%
\bibitem [{\citenamefont {Roy}\ and\ \citenamefont
  {Venugopalan}(2019{\natexlab{a}})}]{Roy:2019hwr}%
  \BibitemOpen
  \bibfield  {author} {\bibinfo {author} {\bibfnamefont {Kaushik}\ \bibnamefont
  {Roy}}\ and\ \bibinfo {author} {\bibfnamefont {Raju}\ \bibnamefont
  {Venugopalan}},\ }\bibfield  {title} {\enquote {\bibinfo {title} {{NLO impact
  factor for inclusive photon$+$dijet production in $e+A$ DIS at small $x$}},}\
  }\href@noop {} {\  (\bibinfo {year} {2019}{\natexlab{a}})},\ \Eprint
  {http://arxiv.org/abs/1911.04530} {arXiv:1911.04530 [hep-ph]} \BibitemShut
  {NoStop}%
\bibitem [{\citenamefont {Roy}\ and\ \citenamefont
  {Venugopalan}(2019{\natexlab{b}})}]{Roy:2019cux}%
  \BibitemOpen
  \bibfield  {author} {\bibinfo {author} {\bibfnamefont {Kaushik}\ \bibnamefont
  {Roy}}\ and\ \bibinfo {author} {\bibfnamefont {Raju}\ \bibnamefont
  {Venugopalan}},\ }\bibfield  {title} {\enquote {\bibinfo {title} {{Extracting
  many-body correlators of saturated gluons with precision from inclusive
  photon$+$dijet final states in deeply inelastic scattering}},}\ }\href@noop
  {} {\  (\bibinfo {year} {2019}{\natexlab{b}})},\ \Eprint
  {http://arxiv.org/abs/1911.04519} {arXiv:1911.04519 [hep-ph]} \BibitemShut
  {NoStop}%
\bibitem [{\citenamefont {Lappi}\ and\ \citenamefont
  {Schlichting}(2018)}]{Lappi:2017skr}%
  \BibitemOpen
  \bibfield  {author} {\bibinfo {author} {\bibfnamefont {Tuomas}\ \bibnamefont
  {Lappi}}\ and\ \bibinfo {author} {\bibfnamefont {Sören}\ \bibnamefont
  {Schlichting}},\ }\bibfield  {title} {\enquote {\bibinfo {title} {{Linearly
  polarized gluons and axial charge fluctuations in the Glasma}},}\ }\href
  {\doibase 10.1103/PhysRevD.97.034034} {\bibfield  {journal} {\bibinfo
  {journal} {Phys. Rev.}\ }\textbf {\bibinfo {volume} {D97}},\ \bibinfo {pages}
  {034034} (\bibinfo {year} {2018})},\ \Eprint
  {http://arxiv.org/abs/1708.08625} {arXiv:1708.08625 [hep-ph]} \BibitemShut
  {NoStop}%
\end{thebibliography}%

\end{document}